\documentclass[%
 reprint,
 amsmath,amssymb,
 aps,
]{revtex4-2}

\usepackage{graphicx}
\usepackage{dcolumn}
\usepackage{bm}
\usepackage{cancel}
\usepackage{graphicx}
\usepackage{subfigure}
\usepackage{hyperref}
\usepackage{appendix}

\begin{document}

\preprint{APS/123-QED}

\title{Localization and Topology in Noncentrosymmetric Superconductors with Disorder }

\author{Jinkun Wang$^{1,2}$  }
\author{Sigma-Jun Lu$^{1}$}
\author{Mei Xiang$^{3}$}
\author{Wu-Ming Liu$^{1,2,}$}%
 \email{wliu@iphy.ac.cn}
\affiliation{
$^1$Beijing National Laboratory for Condensed Matter Physics, Institute of Physics, Chinese Academy of Sciences, Beijing 100190, China   \\
$^2$School of Physical Sciences, University of Chinese Academy of Sciences, Beijing 100190, China  \\
$^3$School of Physics and Electronic Engineering, Xinjiang Normal University, Xinjiang 830054, China
}%

\date{\today}

\begin{abstract}

The celebrated Kitaev chain reveals a captivating phase diagram in the presence of various disorders, encompassing multifractal states and topological Anderson phases. In this work, we investigate the localization and topological properties of a dimerized topological noncentrosymmetric superconductor (NCS) under quasiperiodic and Anderson disorders. Using both global and local characterization methods, we identify energy-dependent transitions from ergodic to multifractal and localized states. Extended multifractal regimes emerge from the competition between dimerization, NCS order, and quasiperiodic modulation. This interplay causes localization to occur preferentially in different energy bands depending on the disorder strength, with the lowest bands exhibiting the highest sensitivity to parameter variations. We employ the real-space polarization method to compute the $\mathbb{Z}_2$ topological invariant, revealing alternating topological and trivial phases as the quasiperiodic potential increases, a behavior distinct from the typical topological Anderson phase diagram. Additionally, the topological states show remarkable robustness against Anderson disorder, providing new insights into topological phase stability in non-centrosymmetric systems. Finally, we propose a feasible experimental scheme based on superconducting Josephson junctions, where NCS-like behavior can be engineered via spatially modulated supercurrents. Our findings highlight the distinct roles of different disorder types in shaping localization and topology, providing insight into the engineering of Majorana zero modes and offering profound implications for topological quantum encryption schemes.

\end{abstract}

\maketitle


\section{Introduction}

Superconductors with intrinsic topological properties have attracted significant attention due to their remarkable and intriguing phenomena \cite{RevModPhys.82.3045, RevModPhys.83.1057, annurev:/content/journals/10.1146/annurev-conmatphys-030212-184337}. Understanding the interplay between topology and superconductivity in these systems is crucial for both fundamental physics and technological applications. The unconventional topological superconductor (TSC) aids in understanding the crucial properties of quantum materials, with the Majorana zero modes (MZMs) hosted in TSCs being considered ideal candidates for use as quantum qubits in fault-tolerant quantum computation \cite{PhysRevLett.94.166802, RevModPhys.80.1083, PhysRevLett.105.177002, PhysRevB.82.214509, doi:10.1126/science.1222360, PhysRevB.104.214502, PhysRevB.104.205131}. Recently, superconductors lacking inversion symmetry, known as noncentrosymmetric superconductors (NCSs), have been extensively researched since the discovery of the heavy-fermion superconductor CePt$_3$Si \cite{PhysRevLett.92.027003, PhysRevB.92.125137, WOS:000827270800010, JPSJ.76.051009, PhysRevB.84.184533, Smidman_2017}. In CePt$_3$Si, the inversion symmetry breaking gives rise to strong spin-orbit coupling, lifting spin degeneracy at the Fermi surface \cite{PhysRevLett.105.097002, PhysRevLett.105.217001, PhysRevB.98.245118, PhysRevLett.124.207001, PhysRevB.105.094523, PhysRevB.107.L041101}.

A recent study shows that external field–induced supercurrents can break both inversion and time-reversal symmetry, driving superconductors into NCS state and leading to crescent-shaped Fermi surfaces in the quasiparticle spectrum \cite{doi:10.1126/science.abf1077}. Subsequent predictions indicate that in such unconventional topological gapless or gapped NCSs, MZMs appear as either flat dispersion bands or vortex-bound modes \cite{PhysRevB.83.224511, PhysRevLett.112.067002, PhysRevMaterials.4.041801, WOS:000319807900001, Schnyder_2015, PhysRevB.109.085139}. This gives rise to a novel type of gapless superconductivity, in which MZMs can be effectively manipulated in spin-helical systems \cite{ PhysRevB.85.020502, PhysRevB.99.245416}. However, the stability and localization behavior of MZMs in disordered NCS systems remain unexplored, which is the focus of this work.

One primary issue in topological quantum systems research is the spatially varying potential, as it is an indispensable element and intrinsic delocalization-localization physics \cite{sau2012realizing, PhysRevB.98.024205, PhysRevB.105.064502, PhysRevA.106.052216}. The random potential and incommensurate potential arise in a vast range of fields due to the experimental accessibility, such as in optical \cite{PhysRevLett.122.110404, PhysRevLett.103.013901}, photonic \cite{ PhysRevB.91.064201}, cavity polariton \cite{goblot2020emergence} and moire lattices \cite{balents2020superconductivity}.  In contrast to the periodic potential which entails Bloch states, the random potential induces a complete Anderson localization of all eigenstates \cite{PhysRevB.107.014202}. Meanwhile, the incommensurate potential, commonly known as a quasiperiodic potential, lies between the completely periodic and random potential regimes \cite{PhysRevLett.126.106803}. The Aubry-Andre-Harper (AAH) model extensively studies the emergence of unique critical phases between the extended and localized phases within this quasiperiodic potential \cite{PGHarper_1955, PhysRevB.105.214203}. These critical phases are characterized by the coexistence of phases with distinct boundaries and mobility edges, or by mixed phases without clear boundaries \cite{PhysRevLett.125.196604, PhysRevB.91.014108, PhysRevB.101.014205}. All eigenstates at the critical phase are locally scale-invariant with several noninteger fractal dimensions, i,e., the multifractal states \cite{PhysRevLett.123.025301, PhysRevB.106.024204, 10.21468/SciPostPhys.12.1.027}. These unconventional multifractal states bring new opportunities in different physics branches such as nonergodic physics, Anderson localization, and transport properties \cite{RevModPhys.91.021001, PhysRevB.82.174411, PhysRevLett.124.200602}. Furthermore, some studies have reported different kinds of mobility edges between the multifractal and extended (or localized) phases \cite{10.21468/SciPostPhys.12.1.027}. 

Another major concern is the interplay of topology and disorder. Pioneering studies have investigated the existence of MZMs in disordered topological superconductors, arguing that a finite amount of superconductivity is required to drive the system into a topological phase \cite{PhysRevB.63.224204, doi:10.7566/JPSJ.86.114707, PhysRevB.105.245144}. Additionally, disorder-induced Andreev-bound states can mimic most signatures of MZMs \cite{PhysRevResearch.2.013377}. It is known that topological states survive in the presence of weak disorder, but topological phases perish in the strong disorder limit. Both random and incommensurate potentials can drive a trivial phase to a topological phase, leading to what is known as the topological Anderson insulator (TAI) \cite{PhysRevB.100.144202, PhysRevLett.115.076601}. Topological charge pumping can be realized in the paradigmatic one-dimensional quasiperiodic model and its various generalizations \cite{roati2008anderson, PhysRevB.88.054204, PhysRevLett.109.106402, PhysRevLett.110.076403}. Recent experiments have observed topological phases with critically localized bulk states in quasiperiodic lattices and non-quantized pumping induced by quasiperiodic disorder \cite{XIAO20212175, nakajima2021competition}. This could bring novel insights into the mechanisms and gap characteristics of TAIs, such as the bulk gap or mobility edges in these phases \cite{PhysRevLett.103.196805, PhysRevB.85.035107}. Furthermore, the TAI phases induced by the quasiperiodic disorder can be accompanied by different localization properties of bulk states, which has garnered our significant attention \cite{PhysRevA.105.063327, PhysRevB.106.224505, PhysRevB.91.115415}.  

Both centrosymmetric and noncentrosymmetric superconductors with disorder potential lead to interesting extended multifractal regions and TAI phases, which are difficult to characterize and whose critical properties are still not well understood. In particular, the one-dimensional dimerized topological superconductors offer remarkable and intriguing phenomena that aid in understanding the crucial properties of TSCs \cite{PhysRevB.90.014505, PhysRevB.96.205428, PhysRevB.96.121105, PhysRevB.109.214306}, however, the combination of disorder and sublattice symmetry breaking has not yet been thoroughly explored. This dimerized model hybridizes the one-dimensional Su–Schrieffer–Heeger (SSH) chain and the Kitaev chain and could be realized through practical topolectrical circuits \cite{WOS:000785680000001}. The periodic alternations of the hopping integrals lead to band inversion and exhibit abundant topological phases \cite{PhysRevB.101.085402, LI2021104837}, with the sublattice index serving as an additional quantum number in the classification of Cooper pairs \cite{PhysRevB.101.214507}. More interestingly, in the TSC with quasiperiodic or non-Hermitian disorder, various anomalous mobility edges emerge in the multifractal phase due to the competition between dimerization and disorder effects \cite{PhysRevB.107.014202, PhysRevB.103.224207}. Of particular relevance to our study, the topological properties of this hybrid system are examined \cite{PhysRevB.100.205302}. The system inherently possesses a unique particle-hole symmetry reminiscent of the SSH model, facilitating the modulation of the number of MZMs \cite{PhysRevB.90.014505, PhysRevB.101.085402}. This presents a unique opportunity to design quantum devices and observe the stability of the MZMs which requires conflicting conditions.

In this work, we investigate the interplay between topology and localization in dimerized noncentrosymmetric topological superconductors under quasiperiodic and Anderson disorders. We employ global and local characterization methods to identify energy-dependent transitions from ergodic to multifractal and localized states. Our results reveal extended multifractal phases arising from the competition between dimerization, NCS order, and disorder effects. In particular, we analyze the system's topological properties using the real-space polarization method to compute the $\mathbb{Z}_2$ topological invariant, which serves as a direct indicator of the presence of MZMs. Our study further demonstrates that quasiperiodic disorder can induce alternating topological and trivial phases, a behavior distinct from conventional topological Anderson insulators. These findings provide valuable insights into the mechanisms governing localization and topology in disordered superconductors, particularly the role of noncentrosymmetry in shaping the phase diagram.

The remainder of this paper is organized as follows. Sec. \ref{Model} describes the model and the symmetry in the clean case. In Sec. \ref{Methods} we present the methods used in this study, including detailed descriptions of the disorder models and computational techniques for both localization and topological properties. Sec. \ref{noncentrotsc} discusses the disorder effects in the time reversal and inversion symmetry breaking noncentrosymmetric topological superconductors. Finally, in Sec. \ref{experimen}, we have designed an achievable experimental setup. A brief conclusion is presented in Sec. \ref{conclusion}.

\section{\label{Model} Dimerized Kitaev chain without inversion symmetry}

\begin{figure}[tbp]
\centering
\includegraphics[scale=0.3]{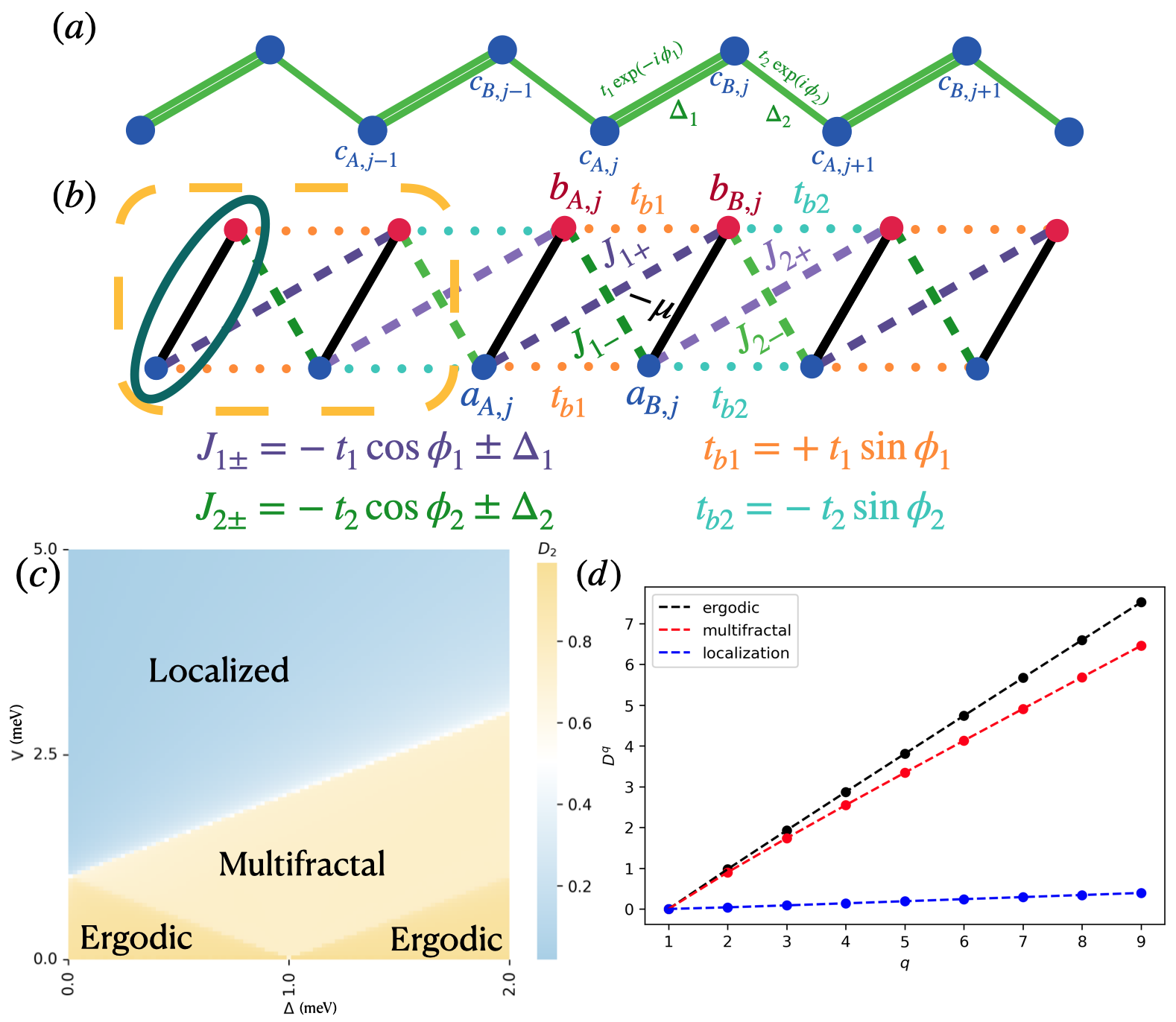}
\caption{ The dimerized topological noncentrosymmetric superconductivity model in (a) complex fermion and (b) Majorana representation. (c) Phase diagram of Kitaev chain under quasiperiodic potential, calculated by mean fractal dimension $D_2$. (d) $D_q$ as a function of $q$, for the ergodic state ($V=0$), $D_q$ maintains a constant slope $1$, corresponding to a uniform fractal dimension; for the localization state ($V=3$), $D_q$ exhibits a slope $0$, indicating a zero fractal dimension; for the multifractal state ($V=0.5$), $D_q$ displays a varying slope and fractal dimension as $q$ increases. Unless otherwise stated, all energy parameters are given in meV. }
\label{phasediabdiqd}
\end{figure}

We start by considering the dimerized Kitaev chain model, which is a prototype model for $p$-wave TSCs, in which the competition between the dimerization and the superconductor pairing order gives rise to a rich phase diagram \cite{PhysRevB.90.014505, WOS:000785680000001}. Here we consider an alternating set of strong and weak bonds with complex hopping amplitudes to break the sublattice symmetry and time-reversal symmetry. As shown in Fig. \ref{phasediabdiqd} (a), the following Bogoliubov-de Gennes (BdG) Hamiltonian is 
\begin{equation}
\begin{aligned}
\mathcal{H}^{\text{BdG}}  &= - \sum_j  \left[ t c^\dagger_{A,j} c_{B,j}  +t^\prime c^\dagger_{B,j}c_{A,j+1} +h.c.  \right]  \\ 
& \quad + \sum_j \left[  \Delta_1  c^\dagger_{A,j} c^\dagger_{B,j}+\Delta_2 c^\dagger_{B,j} c^\dagger_{A,j+1} +h.c. \right]  \\ 
& \quad   - \sum_j \left[  \mu_{1,j} c^\dagger_{A,j} c_{A,j}  +  \mu_{2,j}  c^\dagger_{B,j} c_{B,j} +h.c. \right],  
\end{aligned}
\label{ham0}
\end{equation}
with the two complex different hopping amplitudes $t = t_1 e^{-i\phi_1}$ and $t^\prime = t_2 e^{i\phi_2}$, which could be implemented by an external supercurrent \cite{WOS:000840612900002}. All energy parameters in our tight-binding Hamiltonian are expressed in millielectronvolts (meV). Specifically, we set the real part of the intracell hopping amplitude as the energy unit ($t_1\cos\phi_1 = 1 $ meV), and the imaginary part of the hopping term is considered as the strength of the NCS order.  

So that parameters such as the spatially varying chemical potential $\mu_i = 2V f(i)$ and the superconducting pairings $\Delta_{1,2}$ are naturally defined in meV. Unlike a self-consistent determination from a microscopic mechanism, $\Delta_{1,2}$ in our model are externally set parameters, chosen to satisfy specific gap-closing conditions. This approach reflects their physical origin as proximity-induced superconducting gaps, whose magnitudes are dictated by external constraints rather than a self-consistent pairing interaction. This choice is consistent with experimental observations of superconducting phenomena, where, for example, the proximity-induced superconducting gap is of the order of 0.5 meV \cite{doi:10.1126/science.abf1077}. Without the loss of generality, in the following calculation, we select the two pairings to be real.

 Since we are concerned with the behavior of MZMs, we transfer the Hamiltonian to the Majorana representation, utilizing the relations $c_{A,j} \!=\! (a\!_{A,j}+ib_{A,j})/2$ and $c_{B,j} = (a_{B,j}+ib_{B,j})/2$, which becomes $H \!=\! i  H\!^{Maj}\!/ 2$, where $H\!^{Maj}\! =\! -\mu \sum_j  (  a_{A,j} b_{A,j}  + a_{B,j} b_{B,j} ) -\sum_j  ( J_{1+} a_{A,j} b_{B,j} + J_{1-} a_{B,j} b_{A,j} + J_{2+} a_{B,j} b_{A,j+1} + J_{2-} a_{A,j+1} b_{B,j}   )+ \sum_j  [ t_{b1} ( a_{A,j} a_{B,j} + b_{A,j} b_{B,j} ) - t_{b2} ( a_{B,j} a_{A,j+1 } + b_{B,j} b_{A,j+1}  ) ] $ with the coefficients in each terms $J_{1\pm} = -t_1 \cos \phi_1 \pm \Delta_1$, $J_{2,\pm} = -t_2 \cos \phi_2 \pm \Delta_2$, $t_{b1} = t_1 \sin\phi_1$, $t_{b2} = t_2 \sin \phi_2$. We draw the Majorana representation in Fig. \ref{phasediabdiqd} (b). Compared to the original Kitaev chain, we find the complex hopping terms hybridize the two Majorana chains and connect the isolated MZMs in the boundary. 
 
 In the beginning case of the clean limit, $f(i) =1$, we entail the Bloch theory and employ the Fourier transformation in each site, $c_{\sigma , j} = \frac{1}{\sqrt{N}}\sum_{k_j} \exp {(ik_j r_j)} c_{\sigma , k_j}$, we write our Hamiltonian in momentum space  $\mathcal{H} (k) = \mathcal{H}_0 (k) - \mathcal{H}_{i}  (k)$, specifically
\begin{equation}
\label{hamkt}
\mathcal{H}_0 (k)= - \frac{1}{2}  \tau_z \mu  + \sum_{j = 1,2} M_j (k) \tau_z \sigma_j  + \sum_{j = 1,2} N_j (k) \tau_0 \sigma_j ,
\end{equation}
and the superconductor pairings,
\begin{equation}
\mathcal{H}_{i} (k) \!=\! \frac{1}{2} \! \left[ \left(  \Delta_1 \! - \! \Delta_2 \! \right) \cos k \tau_y  \sigma_y \! + \! \left( \Delta_1 \! + \! \Delta_2 \! \right) \sin k \tau_x \!  \sigma_y  \right]\! ,
\end{equation}
where the $\tau_j$  and $\sigma_j$ are the particle-hole space and sublattice space respectively, the coefficient $M_j (k) = \left(-C_+ \cos k, \> C_- \sin k \right) $,  $ C_\pm = t_1 \cos \phi_1 \pm t_2 \cos \phi_2$ and $N_j (k) = \left( S_+ \sin k, \> S_- \cos k  \right) $, $S_\pm = t_1 \sin \phi_1 \pm t_2 \sin \phi_2$, and $\mu = diag \{ \mu_1, \mu_2 \}$ is the chemical potential.

At the case that both of the $t$ and $t^\prime $ are real, our system respects time-reversal symmetry $\mathcal{T} = \mathcal{K}$, $[ H, \mathcal{T} ] = 0$, particle-hole symmetry $\mathcal{C} = \tau_x \cdot \mathcal{K}$, $\{ H, \mathcal{C} \} = 0$, where $\mathcal{K}$ denotes the antiunitary operator, and the chiral symmetry $\mathcal{S} = \mathcal{T} \cdot \mathcal{C} = \tau_x$, which belong to class BDI case with $\mathbb{Z}$ topological index according to the Atland-Zirnbauer (AZ) classification \cite{RevModPhys.88.035005, Hu_2019}. When we keep the sublattice symmetry at this time, the system comes back to the original Kitaev chain.

While for the imaginary $t $ or $t^\prime $ situation, it simultaneously breaks time-reversal symmetry and chiral symmetry, only particle-hole symmetry is conserved, it belongs to class D case and the topological index is $\mathbb{Z}_2$ \cite{PhysRevLett.62.2747}. The relation between class D and class BDI, and the corresponding topological invariant are shown in Appendix \ref{classbdid}. It is worth mentioning that the inversion symmetry $Ic_{\alpha, j}I^{-1} = \sigma^x_{\alpha,\alpha^\prime} c_{\alpha, -j}$ is also broken in our class D case. This can be viewed in momentum space $Ic_{k,\alpha}I^{-1} = \sigma^x_{\alpha, \alpha^\prime} c_{-k,\alpha^\prime}$. For the inversion symmetry invariant Hamiltonian $IHI^{-1} = H$, i.e. 
\begin{equation}
\begin{aligned}
IHI^{-1} &= \sum_k \sigma^x_{\theta,\alpha} c^\dagger_{-k,\theta} I h_{\alpha,\beta} (k) I^{-1} \sigma_{\beta,\delta} c_{-k,\delta} \\ &= \sum_k c^\dagger_{k,\theta} \sigma^x_{\theta, \alpha} I h_{\alpha, \beta} (-k) I^{-1} \sigma_{\beta, \delta} c_{k,\delta} \> ,
\end{aligned}
\end{equation}
for our system, the inversion symmetry invariant means $h(k) = \tau_0 \otimes \sigma_x h(-k)  \tau_0 \otimes \sigma_x$. Judging by the discussion above, when $\phi_1, \phi_2 \neq 0$, the inversion symmetry is breaking, which means our system enters into topological NCS states. 
 
For the inversion symmetry lacking NCS model without sublattice and time-reversal symmetry, we mainly focus on the gap close point at $k =0$ or $k = \pm \pi/2$, since it usually relates to the topological phase transition. Combine with the determinant of Hamiltonian at this condition $\det H \propto \sin^2 k$ or $\det H \propto 1- \sin^2 k$. We could solve the gapless condition 
\begin{equation}
\left| t \mp t^\prime \right|^2 = \left| \Delta_1 \pm \Delta_2 \right|^2 + \mu_1 \mu_2,
\label{BICcondi}
\end{equation}
which usually distinguish different topological states. 

We are interested in the combination of NCS order and different disorder cases, which include Anderson disorder and AAH modulation. The Anderson disorder is encoded in chemical potential and characterized by a typical on-site disorder, which has uncorrelated uniform distribution $\mu_j = \bar{\mu_j}+ [ -\lambda, \lambda]$ and constant mean $\bar{\mu} \ = \langle \bar{\mu_j} \rangle$ in the pristine case. Similarly, the on-site quasiperiodic modulation is \cite{PhysRevB.107.014202, PhysRevB.106.024204, PhysRevA.105.063327, PhysRevB.106.224505}: 
 \begin{equation}
 \label{qdpocp}
 \mu_i = 2V \cos (2\pi \beta i +\phi ) \>, 
 \end{equation}
 where $\beta^{-1} = \frac{\sqrt{5}-1}{2}$ is the inverse of the golden ratio, $\phi$ is the additional phase shift. A nonzero value of the offset parameter is needed to avoid reflection symmetries in the system, which lead to a degenerate spectrum. Otherwise, the offset parameter value does not affect the multifractal and localization properties of the system. The ratio of two consecutive Fibonacci numbers $\frac{F_{m-1}}{F_m} $ converges to the inverse of the golden ratio for $m \to \infty$. By choosing a length $N = F_m$ for a modulation frequency $\beta = \frac{F_{m-1}}{F_m}$, we make sure that the on-site quasiperiodic potential $f(i)$ is incommensurate within our system concerning the lattice periodicity \cite{PhysRevB.106.024204}. Both of the two disorders break the translation symmetry and the momentum $k$ is not a good quantum number anymore.
 
 \section{\label{Methods} Toolbox for Localization and Topology}

We now introduce the toolbox used to characterize the localization behaviors and topological Anderson states. We propose the topological index in real space to avoid the calculation difficulties caused by translation symmetry breaking in finite disorder. 

\subsection{Localization observables}

To verify the various localized quantum states, we first review the pivotal conclusions about these global observables, i.e., the inverse/normalized participation ratio (IPR/NPR) and the mean fractal dimension $D_2$.  The definition of IPR, $I^{q}_{n,x} $ (NPR, $I^{q}_{n,k} $), of the $n$-th eigenstate in real (momentum) space: 
\begin{equation}
I^{q}_{n,\beta}  = \sum^L_{l=1} | \phi_{n,\beta} (l) |^{2q}, \quad \bar{I}^{q}_{n,\beta} = \frac{1}{L} \sum^L_{j=1} I^{q}_{n,\beta},
\end{equation}
 where $\beta = x,\> k$ respectively denote the sum over the real and momentum sites, $ \phi_{n,\beta} (l)$ denotes the probability amplitude of the $n$-th normalized eigenstate at $l$-site in the $\beta$ space, $\bar{I}^{q}_{n, \beta}$, $\beta = x$ (or $\beta = k$), is the average IPR (NPR). To obtain $I^{q}_{n, k} $ we use the discrete Fourier transformation $\phi_{n,k} (l) = \frac{1}{L} \sum^L_{p=1} e^{-i2\pi pl/L} \phi_{n,x} (p)$. In Appendix \ref{InterpreteIPRNPR}, we illustrate how these observables distinguish the ergodic, multifractal, and localized states. 

Although the IPR and NPR have different behaviors at different localized phases, the scaling behavior is an inevitable influence in it and greatly affects the numerical stability. Fortunately, no matter how we change the system size, the scaling of IPR and NPR is fixed, we could define the fractal dimension 
\begin{equation}
D_q(n) = -\lim_{L\to \infty}\frac{\ln I^{q}_{n,x} }{\ln L},
\label{deffracdim}
\end{equation}
 for each eigenstates $n$ and the mean fractal dimension 
\begin{equation}
\bar{D}_q = - \frac{1}{2N} \sum_{n=1}^{2N} D_q(n),
\label{avefracdim}
\end{equation}
as the global observable.We illustrate the use of the global methods by studying the quasiperiodic potential, Eq. (\ref{qdpocp}), manipulate the ordinary Kitaev chain model, which is the Hamiltonian Eq. (\ref{ham0}) with time-reversal symmetry: $t = t^\prime$ is real, and sublattice symmetry: $\Delta_1 = \Delta_2$, $\mu_1 = \mu_2$. In Fig. \ref{phasediabdiqd} (a), we draw the fractal dimension $D_2$ in terms of the superconducting pairing $\Delta$ and the quasiperiodic potential strength $V$. Which reveals clear phase boundaries $V = |\Delta \pm t| $ between the ergodic phase ($D_2 = 1$), localized phase ($D_2 = 0$) and the multifractal phase ($D_2 $ varies between $0$ and $1$) \cite{PhysRevB.93.104504, PhysRevA.105.013315}. The ergodic states, $D_q \propto q$, or localized states, $D_q \propto 0$, can be defined with only one scaling exponent, drawn in Fig. \ref{phasediabdiqd} (b). While the multifractal states, the slope is neither $0$ or $1$, need several fractal dimensions to fully describe its properties and the gradient of $D_q$ depends on $q$  \cite{SALAT2017467}. 

We introduce two types of powerful local methods to characterize the energy-dependent phase transition. First, to exhibit the different localization behavior for each band, we can still rely on the fractal dimension but instead of taking a mean value, we study it for each eigenstate, such as Eq. (\ref{deffracdim}). In the rest of this work, we mainly focus on the $2$-fractal dimension $D_2$ since it is equivalent to the usual box-counting fractal dimension and gives us information on how the probability distribution of a state fills the space without requiring a finite size extrapolation to classify the different types of states \cite{PhysRevB.103.075124}.

Another observable that could help us to distinguish the localized, ergodic, and multifractal region is the energy level spacing \cite{PhysRevLett.123.025301}, which relies exclusively on the distribution of the eigenenergies $E_n$, instead of the eigenstates. In particular, we compute the even-odd (odd-even) spacing 
\begin{equation}
\begin{aligned}
S^{e-o} = E_{2n} -E_{2n-1},  \quad 
S^{o-e} = E_{2n+1} -E_{2n},
\label{defels}
\end{aligned}
\end{equation}
where $E_n$ are the energy levels for each eigenstate. In our system, we only consider the positive energies and $E_1$ is the minimum positive energy. For the extended states, we expect the eigenenergies to be doubly degenerate, which means $S^{e-o}_n = 0$ and $S^{o-e}_n \neq 0$, a gap should be observed between them. 

One important property of the mobility edge is the blocklike structure. In general, high-energy states are more sensitive to the quasiperiodic potential, and the competition between kinetic and potential energy makes them more susceptible to localization effects. The modulation of quasiperiodic potential separates the different bands by $n/N = \beta^s$ fillings, where $\beta$ is the modulation frequency and $s$ belongs to the integer, $n$ is the number of states below the mobility edge, $N$ is the total number of states or the length of the system. In Appendix \ref{centrotsc}, we apply these methods into the centrosymmetric dimerized Kitaev chain model and find the dimerization order will disturb the blocklike structure and bring unique anomalous mobility edge.

\subsection{Real-space topological invariant}

The system transitions from a topologically nontrivial to a trivial phase under large disorder, but remains robust in the nontrivial phase for weak disorder. To distinguish these phases, a real-space topological invariant is needed, as disorder breaks translational symmetry. Since our system is time-reversal breaking, we mainly focus on the first $\mathbb{Z}_2$ index for class D
\begin{equation}
\gamma = \frac{i}{2\pi} \int^{2\pi}_0 dk \sum_{\alpha \in occ} \langle u_\alpha (k) \left| \partial_k \right| u_\alpha (k) \rangle.
\end{equation}
which is the standard electric polarization, $| u_\alpha (k) \rangle$ is the eigenstates of occupied bands. To measure the displacement of the electron density in real space, we could calculate  the Wannier center within the help of projected position operator $X_\mathcal{P} = \mathcal{P}X\mathcal{P}$, whose imaginary part of eigenvalues corresponds to the center of mass. The position operator $X$ in finite system indicate the state who localize at position $R$, 
\begin{equation}
X = \sum_R e^{\frac{2\pi R}{L}i} | R\rangle \langle R|, 
\end{equation}
with imposed periodic boundary condition. Where L is the total number of unit cells in the system and the exponent ensuring periodic boundary conditions. Since we only consider the occupied bands in projection operator
\begin{equation}
P = \sum^{N_{occ}}_{n}| \psi_n \rangle \langle \psi_n |,
\end{equation}
$| \psi_n \rangle$ is the eigenstate of the $n$-th occupied band and the dimension of projector operator is same to the number of lattice site $L$. Hence, we could represent the polarization in real-space as \cite{PhysRevLett.113.046802}:
\begin{equation}
\gamma = \sum^L_l \left( \frac{1}{2\pi} \text{Im} \log \xi_l - \frac{l}{L} \right) \mod 1,
\label{rspolar}
\end{equation}
where $\left\{ \xi_l \right\}$ is the eigenvalues of $X_{\mathcal{P}}$, and $l$ is the eigenstate index of the projected position operator $X_P $, with a range of values equal to the lattice size. This formula is fitted for the disorder system since the Wilson loop method is not applicable without translation symmetry. The relation between the polarization and winding number is discussed in Appendix \ref{topoinvar}.

\begin{figure}[tbp]
\centering
\includegraphics[scale=0.36]{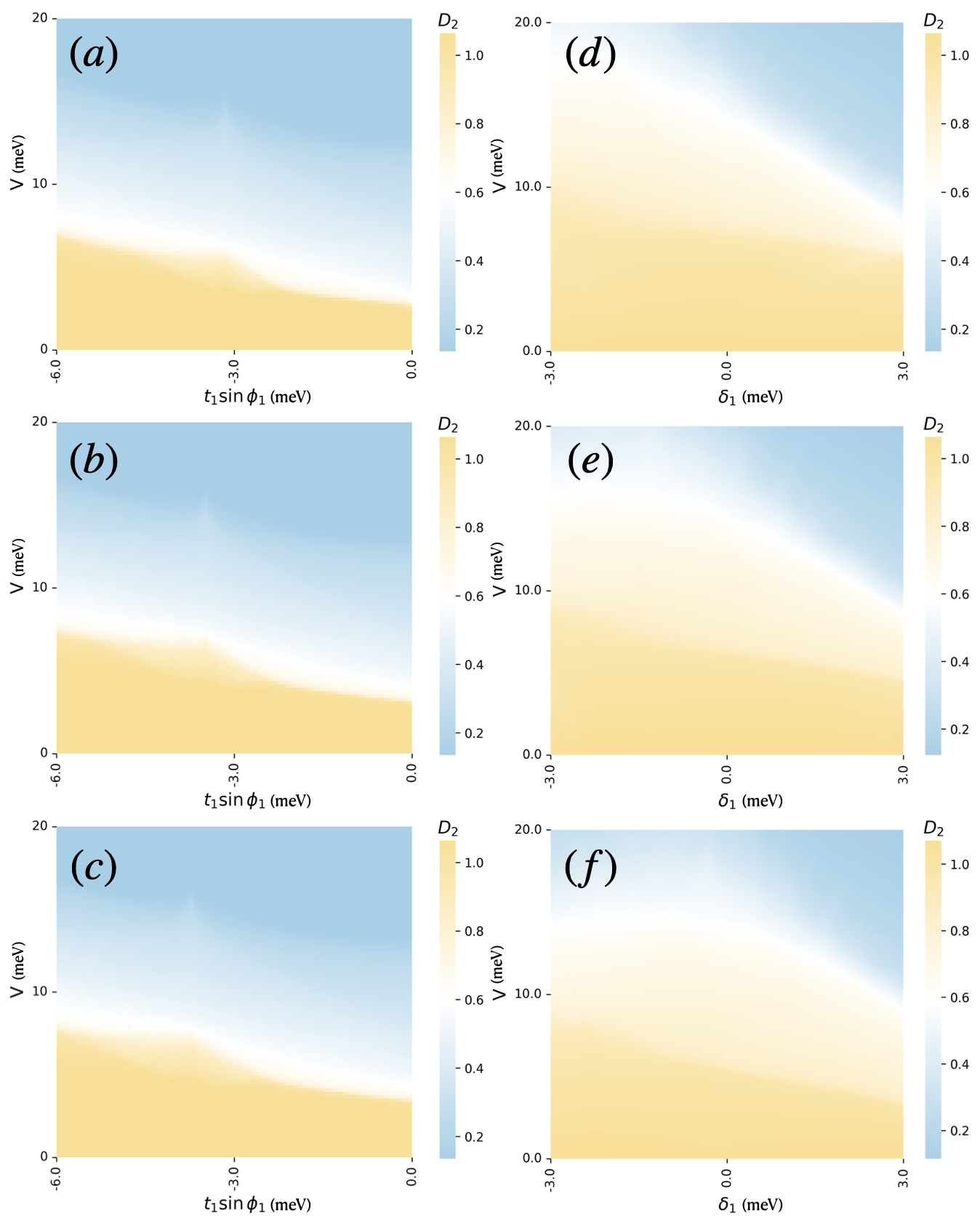}
\caption{Localization Phase diagram of the model with time-reversal breaking NCS states. (a)-(c) Mean fractal dimension $D_2$ for (a) $t_2 \sin \phi_2 = 1.73$, (b) $t_2 \sin \phi_2 = 2.31$, (c) $t_2 \sin \phi_2 = 2.65$, the system stays on gapless all the time. (d)-(f) We open the systems gap at $t_1 \sin \phi_1 = -5.9$ and $t_2 \sin\phi_2 = 1.5$ and calculate the mean fractal dimension $D_2$ for (d) $\delta_2 = -2.63$, (e) $\delta_2 = -1.31$, (f) $\delta_2 = -0.38$. For all the plots we consider $t_1\cos \phi_1 = 4.6$, $t_2\cos \phi_2 = -5.3$ and $N = 1597$. }
\label{fracdimD}
\end{figure}

\section{\label{noncentrotsc}The Non-Centrosymmetric topological superconductors }

In this section, we focus on the localization and topological properties of noncentrosymmetric topological superconductors. While previous studies have primarily addressed time-reversal and inversion-symmetric cases, a detailed discussion of the inversion-symmetric dimerized Kitaev chain under disorder can be found in Appendix \ref{centrotsc}. Here, we shift our focus to the unique class D topological superconductors, where the breaking of inversion symmetry leads to distinct localization and topological features. Specifically, we consider the Hamiltonian Eq. (\ref{ham0}) and analyze how disorder strength, hopping amplitudes, superconducting pairings, and chemical potentials affect the localization and topological phase transitions.

We first examine different regions of the localization phase diagrams of our model using the mean fractal dimension $\bar{D}_2$. Next, we investigate energy-dependent transitions by analyzing individual eigenstates, aiming to identify possible anomalous mobility edges within multifractal regions. Finally, we explore the unique topological behaviors induced by various types of disorder potentials.

 \subsection{  Localization }
 
 We first characterize different regions of the global nature of our system for different cuts in the parameter space. We mainly focus on the gap close and reopen processes which usually relate to the phase transition behaviors. According to the gap close condition solved in Eq. (\ref{BICcondi}), we could decompose the hopping into real and complex parts respectively $t = t_1 \cos \phi_1 + i t_1 \sin \phi_1$ and $t^\prime = t_2 \cos \phi_2 + i t_2 \sin \phi_2$. At the gapless at $k = 0$, when we fix the two real part $ t_1 \cos \phi_1 $ and $t_2 \cos \phi_2$, then we select the parameters of two imaginary parts $t_1 \sin \phi_1  $ and $t_2 \sin \phi_2$  to change freely, in the meantime the two pairings $\Delta_{1,2}$ are determined based on the gapless condition when other parameters changes, $\Delta_1 = -\Delta_2 = \Delta_c$, where $\Delta_c = |t+ t^\prime|^2/2$. We choose the chemical potential $\mu = 0$ since it usually tends to create isolated MZMs. 
 
Fig. \ref{fracdimD} (a)-(c) display the mean fractal dimension $\bar{D}_2$ for the gapless case as a function of the quasiperiodic potential $V$ and $t_1 \sin \phi_1$, with fixed values of $t_2 \sin \phi_2 = 1.73$ (a), $t_2 \sin \phi_2 = 2.31$ (b), and $t_2 \sin \phi_2 = 2.65$ (c). The pairing terms are determined under the gapless condition. As $V$ increases, the system undergoes a transition from ergodic to localized states for any finite value of $t_2 \sin \phi_2$. However, the phase boundaries between ergodic and intermediate states are not sharply defined throughout; for certain finite values of $t_2 \sin \phi_2$, the phase boundary extends into a bubble of multifractal states. As the wave function gradually transitions into the localized phase, a distinct phase boundary becomes indiscernible. This indicates that the transition is energy-dependent, meaning not all eigenstates exhibit the same localization properties, which we will further analyze. Notably, the ergodic state remains more robust, and the multifractal state bubble gradually expands as $t_2 \sin \phi_2$ increases.
 
In Fig. \ref{fracdimD} (d)-(f), we focus on the gapped case by adjusting the two superconducting pairings from the gapless condition determined pairing $\Delta_c$ to $\Delta_1 = \Delta_c - \delta_1$ and $\Delta_2 = -\Delta_c + \delta_2$. We use the gapless parameters $t_1 \sin \phi_1 = -5.9$ and $t_2 \sin \phi_2 = 1.5$, varying $\delta_1$ freely while fixing $\delta_2$ at values of $-2.62$ (d), $-1.31$ (e), and -0.37 (f). In contrast to the gapless case, distinct boundaries between different phases emerge, and the multifractal phase disperses over a larger region, further suggesting that transition points may differ across eigenstates. Additionally, this energy-dependent phase behavior becomes more pronounced with increasing $t_2 \sin \phi_2$, complicating the study of multifractal intermediate regions using global characterization methods. In the next section, we will apply the local tools from Sec. \ref{Methods} B to examine individual eigenstates and investigate the phase transitions in greater detail.
 
 \begin{figure}[tbp]
\centering
\includegraphics[scale=0.3]{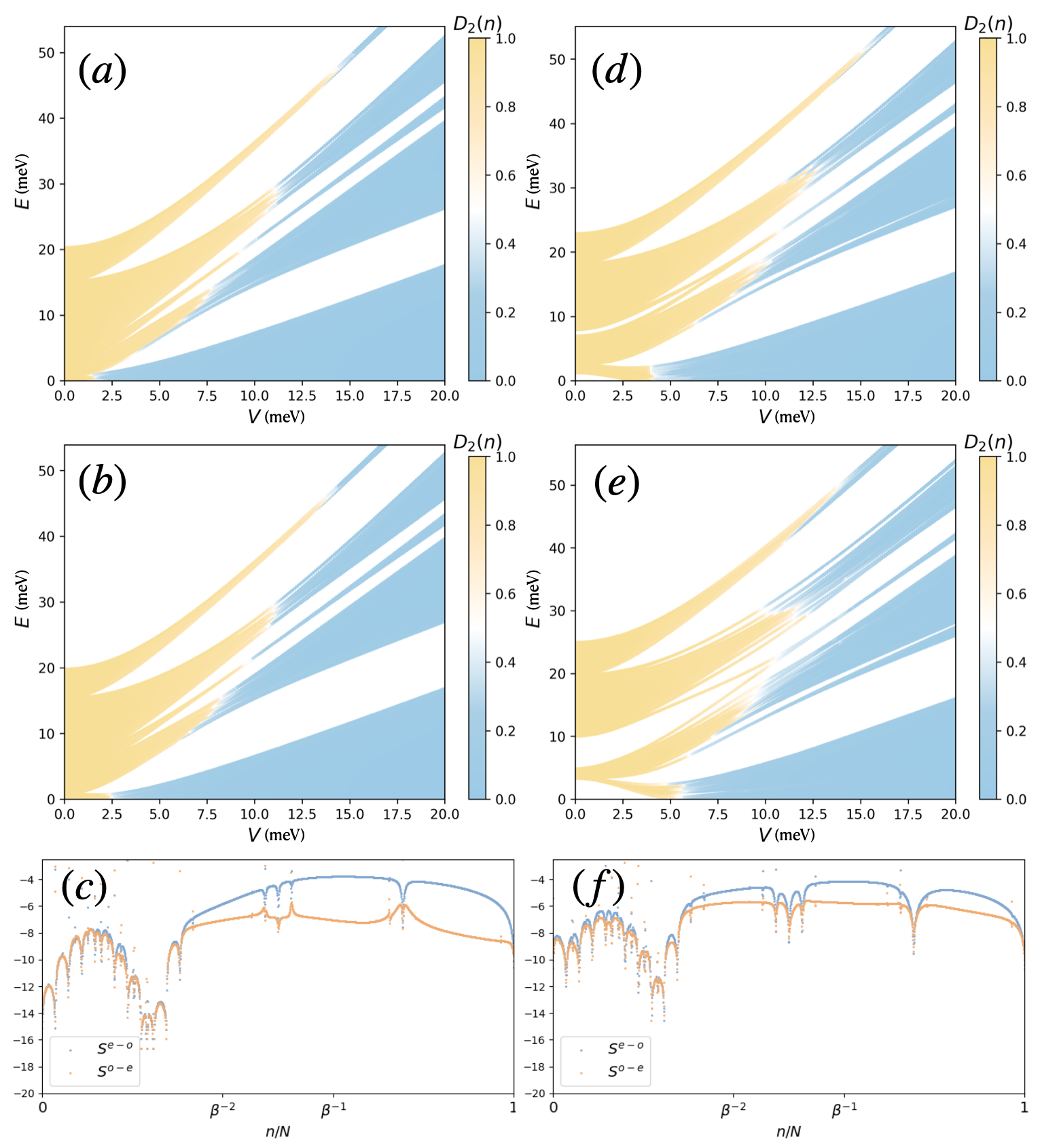}
\caption{The energy-dependent localization phase transitions. The positive energy spectrum for different quasiperiodic potential $V$. The color bars indicate the fractal dimension $D_2(n)$ for each eigenstates corresponds to the energy $E$, computed for the gapless case with, (a) $t_1 \sin\phi_1 = -0.06$, (b) $t_1 \sin \phi_1 = -2.640$; and the gapped case with (d) $\delta_1 = -0.84$, (e) $\delta_1 = -2.94$, band gaps are opened at $t_1 \sin \phi_1 = -5.9$ and $t_2 \sin\phi_2 = 1.5$. The logarithm of even-odd $S^{e-o}_n$ (odd-even $S^{o-e}_n$) energy level spacings in red (black) for  (c) $t_1 \sin \phi_1 = -2.640$,  $V = 2.218$ (f) $\delta_1 = -2.94$, $V = 4$.  Other parameters are selected as $t_1\cos \phi_1 = 4.6$, $t_2\cos \phi_2 = -5.3$, $N = 1597$.
 }
\label{MultiFracD}
\end{figure}

\begin{figure*}[htbp]
\centering
\includegraphics[scale=0.36]{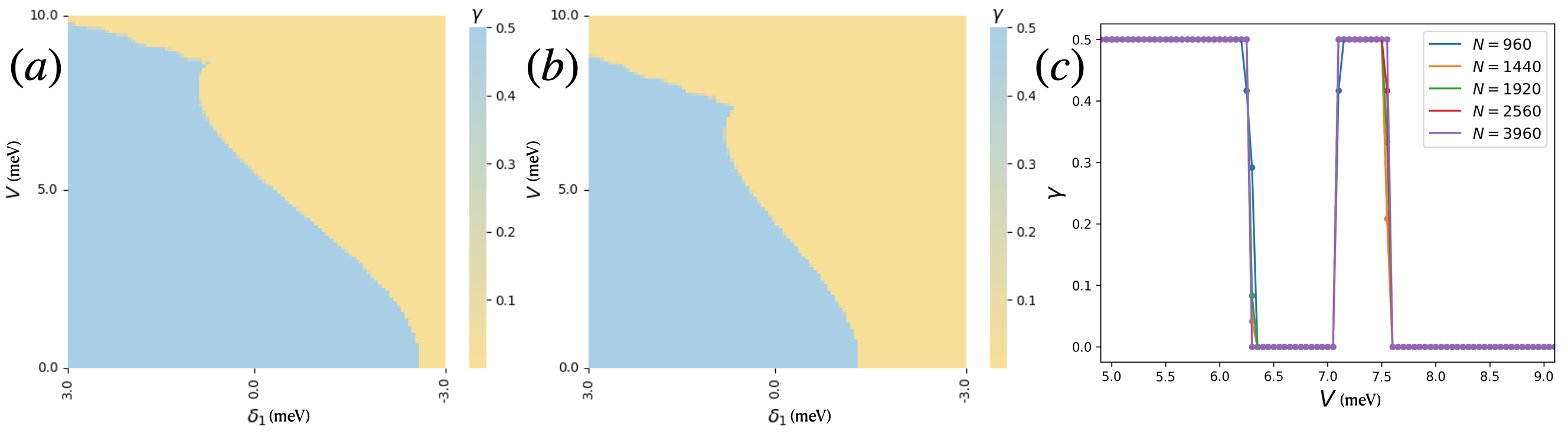}
\caption{ The topological phase diagram for the noncentrosymmetric superconductor, calculated by real-space polarization $\gamma$ as the function of $\delta_1$ and quasiperiodic potential $V$ for the (a) $\delta_2 = -2.62$, (b)  $\delta_2 = -1.31$, we select the system size $N = 1597$ (c) the real-space polarization calculated in  $\delta_2 = -1.31$ with different system length $N = 960$, $  1440$, $ 1920$, $ 2560$, $ 3960$. Other parameters are selected as  $t_1\cos \phi_1 = 4.6$, $t_2\cos \phi_2 = -5.3$, $t_1 \sin \phi_1 = -5.9$ and $t_2 \sin\phi_2 = 1.5$.
}
\label{polarD}
\end{figure*}

To analyze the localization behavior of individual energy bands in greater detail, Fig. \ref{MultiFracD} presents the positive portion of the energy spectrum as a function of the quasiperiodic potential $V$ for two cases: (i) the critical gapless state, with $t_2 \sin \phi_2$ fixed at 2.31 and different values of the NCS order strength, specifically (a) $t_1 \sin \phi_1 = -0.06$ and (b) $t_1 \sin \phi_1 = -4.38$; and (ii) the gapped case, with $\delta_2$ fixed at -0.38 and (d) $\delta_1 = -1.02$, (e) $\delta_1 = 2.94$. In these plots, the color map indicates the fractal dimension $D_2$ of each eigenstate.

Our observations reveal that, as expected, the transition to localization depends on energy across all cases. However, in gapless cases with a weak quasiperiodic potential  $ V $, we do not observe a distinct mobility edge, as would be represented by a clear boundary line or blocklike structure in the energy spectrum. This indicates that the localization transition lacks a straightforward, universal threshold across energy levels, resulting in a more complex and spatially diverse localization behavior than initially anticipated. Notably, we identify a transition bubble region, which functions as a flexible boundary between extended and localized states. Within this bubble, the localization characteristics of the lower energy bands shift rapidly, illustrating a sensitive response to even small changes in the potential, which drives a fluid, energy-dependent transition.

In the gapped case, the larger energy gap between states results in localization occurring more readily, as states are well-separated and less susceptible to the influence of delocalizing factors. This reinforces the general trend that, in both gapless and gapped scenarios, lower energy bands are more prone to localization due to their greater sensitivity to disorder. The observation that localization initiates in the lower bands highlights how energy impacts the response of different states to disorder, with lower bands more readily adopting localized character.

In addition, the localized block structure observed does not strictly scale with the power of $\beta$, suggesting that the localization pattern is not solely determined by potential strength in a simple, linear manner. Instead, this reveals a more intricate relationship between the quasiperiodic potential and the localization properties of the system, hinting at richer, less predictable phase transitions that are likely influenced by the interplay of potential periodicity, energy band structure, and the underlying disorder. Understanding this nuanced relationship offers deeper insights into the localization phenomena and complex phases of quasiperiodic systems.

 \subsection{ Topology  }

\begin{figure*}[tbp]
\centering
\includegraphics[scale=0.38]{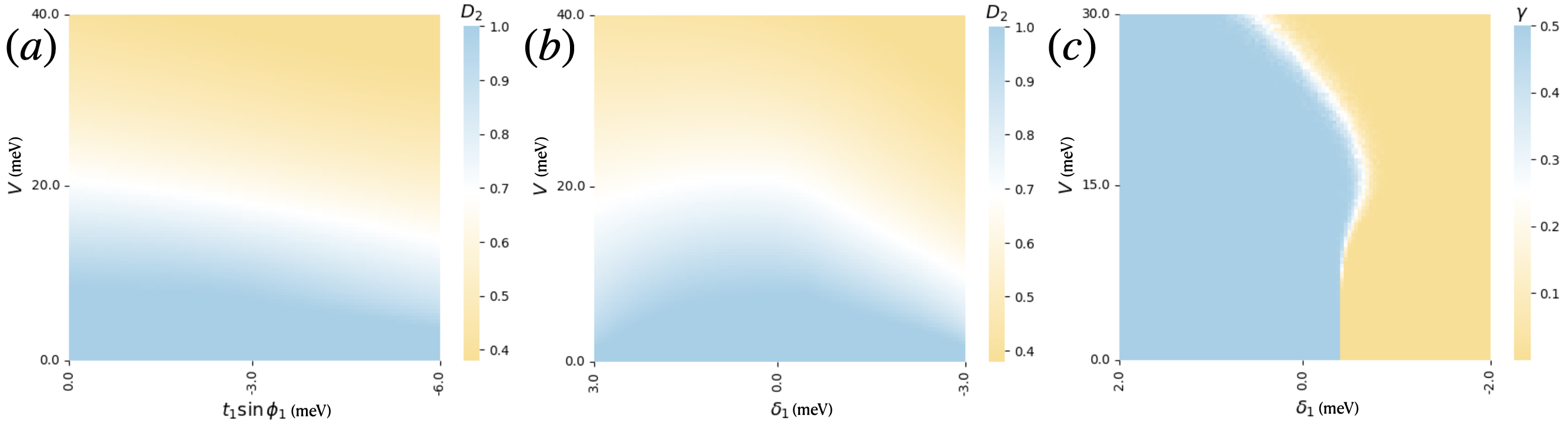}
\caption{ The localization and topological behaviors of noncentrosymmetric superconductors with Anderson disorder. For the gapless case (a) the mean fractal dimension $D_2$ at $t_2 \sin \phi_2 = 2.65$; and the gapped case (b), mean fractal dimension $D_2$ and (c) real-space polarization at  $t_1 \sin \phi_1 = -5.9$ and $t_2 \sin\phi_2 = 1.5$ and $ \delta_2 = -0.37$.  Other parameters are selected as  $t_1\cos \phi_1 = 4.6$, $t_2\cos \phi_2 = -5.3$ and system size $N = 1597$.
}
\label{AndersonLT}
\end{figure*}
 
 In this section, we investigate the topological properties of the MZMs in our model. The polarization, Eq. (\ref{rspolar}), is the only suitable topological invariant in this time-reversal symmetry breaking case, which could distinguish the topological superconductor state and the trivial state in the gap opening situation. We draw the phase diagrams via the polarization calculated in real space $\gamma$ in Fig. \ref{polarD}, where the topological phases are indicated as $\gamma = 1/2$ while the trivial phases with $\gamma = 0$. Similarly, the band gap is opened by the bias of gapless condition determined pairing $\Delta_c$.
 
This disorder-induced topological state is an exciting discovery because it challenges the traditional view that topological phases are solely determined by the intrinsic symmetries of the system, independent of disorder. Here, we observe the topological phase transition point shifting from $\delta_2 \sim 2.8$ to $\delta_2 \sim 0$ as the parameter $\delta_1$ increases, under the limit of $V=0$. In the case of a finite quasiperiodic potential $V$, the area of the topological phase continues to decrease with the increase of $\delta_1$. In contrast to the traditional topological Anderson superconductor, the finite disorder did not generate the topological states from the trivial states in clean limit. However, for the topological phase in the clean case, after crossing the topology-trivial phase boundary, as the increasing quasiperiodic potential $V$, the system reenters into the topological phase. Our results show that the topological Anderson superconductor state does not only appear as a continuous phase once in the phase diagram but can also intersect with a trivial state under limited disorder modulation.
 
Topological superconducting phases host MZMs as zero-energy boundary states protected by bulk topology. In our model, these modes correspond to the topological phase with polarization $\gamma = 1/2$ region. In contrast to conventional topological superconductors, in which disorder typically weakens MZMs, our system exhibits reentrant topological phases where the MZMs remain robust even in the presence of strong disorder. This behavior is clearly visible in the polarization phase diagram, Fig. \ref{polarD}, where MZMs continue to appear in disorder-induced topological regions. This indicates that the interplay between noncentrosymmetry, dimerization, and disorder stabilizes MZMs across a broad parameter range.
  
Furthermore, we perform a finite-size analysis to characterize the topological Anderson phase transition, as shown in Fig. \ref{polarD} (c). In this calculation, we plot the real-space polarization $\gamma$ as a function of the quasiperiodic potential $V$ for various system sizes $L = 960$, $  1440$, $ 1920$, $ 2560$, $ 3960$, with each size represented by a different color. The variation of $\gamma$ with $V$ exhibits a sharp transition from $1/2$, topological phase, to $0$, trivial phase, as $V$ increases. Notably, this transition appears consistent across all system sizes, confirming the robustness of the topological Anderson phase transition. This result aligns perfectly with the conclusions drawn from the phase diagram in Fig. \ref{polarD} (b), which demonstrates that increasing the quasiperiodic disorder leads to alternating regions of trivial and topological phases, forming an intricate, interwoven topological phase diagram. This finite-size scaling analysis confirms that the transitions between topological and trivial phases remain sharp and size-independent, reinforcing the notion that the disorder-modulated topology in our model is a robust physical phenomenon rather than a finite-size artifact. These results provide new insights into the stability of topological superconducting phases in disordered systems and offer a promising pathway for realizing disorder-resilient MZM-based quantum devices.

\subsection{Effect of Anderson disorder}

In the final section, we explore and contrast the effects of Anderson disorder and quasiperiodic potential on the localization and topological properties of our system. To achieve this, in Fig. \ref{AndersonLT},  we examine the localization phase diagrams generated via the mean fractal dimension and the topological phase diagrams derived from real-space polarization, shedding light on how Anderson disorder impacts localization onset and MZMs.

In Fig \ref{AndersonLT} (a) and (b),  our findings reveal a gradual transition into the localized phase, without a clear-cut phase boundary separating ergodic, intermediate, and fully localized states. Although an intermediate regime exists, it is not a multifractal state as seen in quasiperiodic systems. Instead, all wave functions within this intermediate state share similar localized properties, lacking the diversity of localization behavior that typically characterizes multifractal phases. This uniformity indicates that Anderson disorder disrupts extended states more uniformly across energy levels. However, it takes a relatively strong Anderson disorder to fully drive the system into a fully localized state, highlighting the disorder's strength's critical role in pushing the system across the localization threshold.

Notably, we observe that MZMs demonstrate remarkable resilience in the face of Anderson disorder compared to quasiperiodic disorder, drawn in Fig. \ref{AndersonLT} (c). These topological edge states persist even as the disorder strength increases, underscoring their robustness and the protection offered by the system's inherent topological properties. This robustness contrasts with the bulk states, which transition more readily toward localization under increasing disorder. The MZMs’ persistence highlights the stability of topological features in disordered systems, a trait that can be instrumental in quantum information applications where robustness against environmental disorder is crucial.

These two disorders affect localization and topology in distinct ways. In contrast to the quasiperiodic disorder with clear mobility edge and multifractal intermediate states, Anderson disorder gradually localizes states without sharp boundaries, resulting in a more uniform transition. Furthermore, compared with the influence of Anderson disorder, the quasiperiodic disorder's structured, phase-sensitive localization more intricately influences topological phases.

\section{\label{experimen}Experimental Implementation}

\begin{figure}[tbp]
\centering
\includegraphics[scale=0.47]{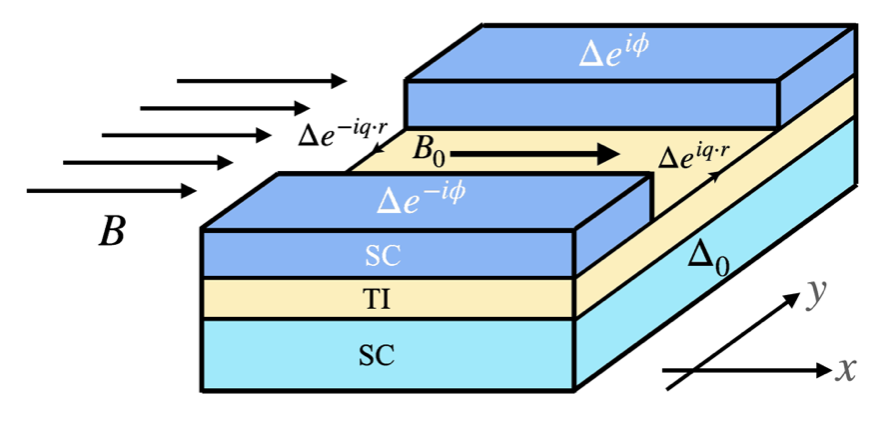}
\caption{  Schematic of a Josephson junction designed to realize a topological NCS and host MZMs. A topological insulator is proximity-coupled to two $s$-wave superconducting leads with a relative phase difference of $2\phi$. An in-plane magnetic field $B$ induces a screening current $q$ along the $y$-direction between the superconducting leads, generating a spatially varying superconducting phase that breaks inversion symmetry. }
\label{ExperPhasTrans}
\end{figure}

 Numerous theoretical and experimental approaches have been explored to detect Kitaev chain physics and MZMs \cite{doi:10.1126/science.aan3670, doi:10.1021/acs.nanolett.7b01728}. An effective strategy is to employ planar Josephson junctions fabricated on the conductive 2D surface of a strong 3D topological insulator (TI), as depicted in Fig. \ref{ExperPhasTrans}. By applying an in-plane Zeeman field $B_0$ and utilizing the robust spin-orbit coupling, the proximity effect from superconducting or ferromagnetic layers can open an energy gap in the surface states \cite{PhysRevX.7.021032, PhysRevLett.126.027001, PhysRevB.103.115423, ruanobservation2024}. This creates a quasi-one-dimensional channel between two superconductors, which carry a relative phase difference $\phi$. Under these conditions, especially when $\phi$ approaches $\pi$, the channel effectively realizes a Kitaev-chain-like topological superconductor, hosting MZMs at its ends. Experimentally, these MZMs can be observed via a zero-bias conductance peak in tunneling spectroscopy at the junction edges, or inferred from a sharp minimum in the critical current as the Zeeman field is tuned, indicating a topological phase transition \cite{PhysRevB.105.054504, PhysRevB.86.214515, PhysRevLett.124.227001}. Alternatively, we can scan the magnetic field distribution on the surface of the TI layer, thereby confirming these MZMs at the edges.

In our planar Josephson junction, superconducting electrodes are interfaced with the topological material, where the superconducting proximity effect induces pairing correlations in the semimetal. When an in-plane magnetic field is applied, it produces a Zeeman effect that shifts the momentum of the spin-momentum-locked topological surface states \cite{doi:10.1126/science.abf1077, PhysRevB.97.115139,WOS:000684834100022}. This momentum shift leads to an additional spatially dependent phase in the superconducting order parameter, so that it takes the form $\Delta (\textbf{r}) = \Delta\exp(i \textbf{q} \cdot \textbf{r})$, where $\textbf{q}$ represents the finite center-of-mass momentum acquired by the Cooper pairs \cite{doi:10.1126/science.abf1077}. After making a gauge transformation and breaking sublattice symmetry, we finally obtain the dimerized topological NCS Hamiltonian Eq. (\ref{ham0}). The detailed mechanism is record in Appendix \ref{SupercurrentindSC}. Although asymmetric screening currents in the superconducting leads contribute slightly, the dominant mechanism for generating finite momentum pairing is this Zeeman-driven shift of the topological surface states, which is further validated by the evolution of the Fraunhofer interference pattern under varying magnetic fields \cite{PhysRevX.7.021032}.

In specific experimental, by leveraging intrinsic lattice variations similar to those used in the tellurium surface study \cite{nakayama2024}, one can engineer a dimerized two-sublattice structure within a lattice. For instance, targeted ion-beam sputtering combined with chemical pre-treatment can induce selective surface reconstruction, leading to site-selective dimerization: some atomic chains in the lattice pair up which forming a dimerized sublattice, while others remain with open ends. This spatial modulation creates an alternating pairing potential analogous to the alternating bonds seen in the one dimensional chains. Such an engineered configuration not only mimics the alternating superconducting gap but also provides a flexible, tunable platform to realize topologically nontrivial phases, including the emergence of MZMs.

Notably, the controllable disorder, such as tuning local doping levels or introducing random variations in coupling parameters, arises systematically modulate the topological phase and MZMs \cite{wangwave2024}. Additionally, boundary conditions and finite-size effects can be leveraged to enhance or suppress localized states.  Our analysis of the size effect, in Fig. \ref{polarD} (c), indicates that when the number of lattice sites reaches $3960$, the process of topological phase transition becomes more pronounced. Considering that the lattice constant in materials is generally on the order of angstroms, we believe that in our experimental setup, the effects of disorder in the system become more significant when the distance between two electrodes exceeds $350$ nm.

Above all, this Josephson junction setup offers a streamlined approach to realizing topological superconductor networks. By adjusting a global phase difference, multiple junctions can be simultaneously tuned into a topological phase without the need for individual parameter adjustments or local probes. This experimental configuration provides a promising avenue for the controlled observation and manipulation of MZMs. Furthermore, the ability to engineer spatially modulated pairing and hopping amplitudes through proximity effects and external supercurrents lays the groundwork for more sophisticated topological quantum devices. This platform can be extended to investigate tunable noncentrosymmetric superconducting phases, study disorder-induced topological transitions, and explore potential applications in robust Majorana-based quantum computation.

\section{\label{conclusion}conclusion}

We conducted a comprehensive study of dimerized one-dimensional topological noncentrosymmetric superconductors, examining their localization and topological properties under various disorder influences. We first reviewed methods for characterizing localization and real-space topological invariants. Building on this, we explored the energy-dependent localization behaviors arising from the effect of quasiperiodic potential, observing diverse multifractal states and distinct mobility edges in gapless systems. As demonstrated in the manuscript, the lack of inversion symmetry makes the system’s energy gap more sensitive to the strength of dimerization and disorder, leading to novel localization behavior and the emergence of multifractal phases in certain regions. In gapless systems, lower-energy bands localized more easily, highlighting a nonlinear relationship between localization and quasiperiodic potential strength. Larger energy gaps resulted in sharper phase boundaries between ergodic, multifractal, and localized states. 

Using real-space polarization, we characterized topological phases under disorder with MZMs emerging in the topological phases and disappearing in the trivial phases. Unlike traditional topological Anderson superconductors, finite disorder did not induce topological transitions but instead led to alternatively phase shifts and evolving boundaries across parameter regimes. Our comparison of Anderson and quasiperiodic disorders showed that MZMs exhibit greater resilience under Anderson disorder, maintaining robust topological properties even at large disorders. Finally, we propose planar Josephson junctions as a feasible experimental scheme, offering a promising platform for realizing our topological NCSs. This work lays the groundwork for exploring localization behaviors and engineering topological superconductors, providing valuable insights for practical quantum devices and advancing topological quantum encryption.

\acknowledgments
This work was supported by National Key Research and Development Program of China under grants No. 2024YFF0726700, 2021YFA1400900, 2021YFA0718300, 2021YFA1402100, NSFC under grants No. 12174461, 12234012, 12334012, 52327808.

\appendix

\section{\label{classbdid} The One-dimensional Class BDI and Class D Topological Superconductors}
Our system contains both class BDI and class D topological superconductor models, through adjusting the parameters in it. The difference between them is the lack of time-reversal and chiral symmetry. In this section, we give a metaphysical characterization of them beyond a specific model.  First, the discrete symmetry, which include parity $P: (t,x) \to (t,-x)$, time-reversal $T: (t,x) \to (-t,x)$ and charge conjugation $C$. In lattice, $P a_p^s P = \eta_a a_{-p}^s$, $Tc_k T^{-1} = c_{-k}$, $C a^s_{p} C = b^s_p$. The symmetry invariant Hamiltonian can be represented by 
 \begin{equation}
 \begin{aligned}
 TH(k)T^{-1} &= H(-k), \\ CH(k) C^{-1} &= -H(-k), \\ S H(k) S^{-1} &= -H(k).
 \end{aligned}
 \label{symcon}
 \end{equation}
Before we discuss the symmetry-protected Hamiltonian, we first review the chiral symmetry. Since $S H(k) S^{-1} = -H(k)$, $S^2 = 1$ and $S| u^\pm_s \rangle = \pm | u^\pm_s \rangle$, which means $\langle u^\alpha_s | S H(k) S^{-1} | u^\beta_s \rangle = - \langle u^\alpha_s | H(k) | u^\beta_s \rangle$, then $(-1)^{\alpha+\beta} \langle u^\alpha_s |  H(k) | u^\beta_s \rangle = - \langle u^\alpha_s | H(k) | u^\beta_s \rangle$. The relations are $\langle u^+_s | H | u^+_s  \rangle = 0$ and $\langle u^-_s | H | u^-_s  \rangle = 0$.

Consider the Hamiltonian in Bloch representation $h(k) = \sum_{i} d_i (k) \sigma_i$, $i = x,y,z$, combine Eq. (\ref{symcon}) the time-reversal symmetry and particle-hole symmetry confine that
\begin{equation}
\begin{aligned}
T h(k) T^{-1} &= \begin{pmatrix} d_x (k), & -d_y(k),  & d_z(k) \end{pmatrix} \\ &= \begin{pmatrix} d_x (-k), & d_y(-k),  & d_z(-k) \end{pmatrix},
\end{aligned}
\end{equation}
and
\begin{equation}
\begin{aligned}
C h(k) C^{-1} &= \begin{pmatrix} d_x (k), & d_y(k),  & -d_z(k) \end{pmatrix} \\ &= \begin{pmatrix} -d_x (-k), & -d_y(-k),  & -d_z(-k) \end{pmatrix}.
\end{aligned}
\end{equation}
 
 In the case which maintains the time-reversal symmetry i.e., class BDI case, the Hamiltonian has to maintain the $T$ and $C$ symmetry simultaneously, which means $d_x =0$, $d_y$ is odd and $d_z$ is even. Thus, we can construct the Hamiltonian with all long-range couplings
 \begin{equation}
 d_y (\phi) = \sum_{n=1} \lambda_n \sin (n \phi ), \quad  d_z (\phi) = \sum_{n=1} \lambda^\prime_n \cos ( n \phi )  .
 \label{hamchirI}
 \end{equation}
The trajectory of $\vec{d} = (d_x, d_y, d_z)$ is confined at the $d_y$-$d_z$ plane.According to the definition of winding number 
\begin{equation}
\nu_1 = \frac{i}{2\pi} \int_k dk \left( -d_y +id_z \right)^{-1} \partial_k \left( -d_y +id_z \right),
\end{equation}
combine with the normalization condition for $\left| u^- \right>$: $d_y \partial_k d_y + d_z \partial_k d_z = 0$, the winding number is 
\begin{equation}
\nu_1 = \frac{i}{2\pi} \int_k \partial_k \left[ \arctan \left( \frac{d_z}{d_y} \right) \right],
\end{equation}
which is the trajectory of the Hamiltonian $\vec{d}$, the topological index is $\mathbb{Z}$.
 
\begin{figure}[htbp]
\centering
\includegraphics[scale=0.28]{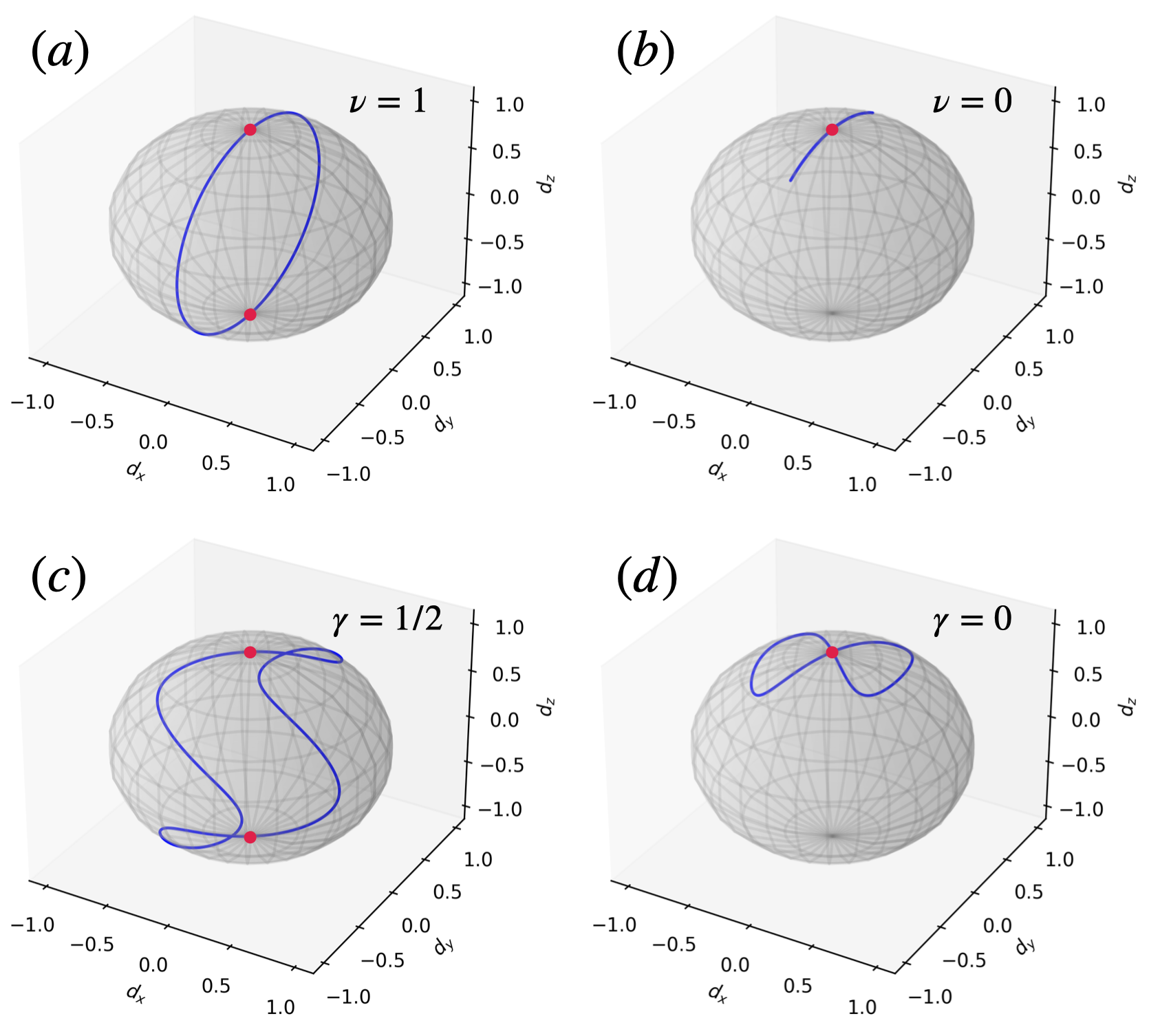}
\caption{ The topological superconductor with time-reversal symmetry cases and the winding number (a) $\nu = 1$ and (b) $\nu =0$. The time-reversal symmetry breaking cases with the polarization (c) $\gamma = 1/2$ and $\gamma = 0$. The parameters are $\lambda_{y1} = \lambda_{z1} =1$ for (a), (c);  $\lambda_{y1} = 1$, $\lambda_{z0} = 2$, $\lambda_{z1} = 1$ for (b), (d) and $\lambda_{x,1} = 0.5$ for (c), (d).}
\label{WinNumdef}
\end{figure}

 While, for the time-reversal symmetry breaking case i.e., class $D$ case, $T =0$ and $C =1$, which means the $d_x$ is odd. The coefficients must satisfy 
  \begin{equation}
  \begin{aligned}
 d_x (\phi) &= \sum_{n=1} \lambda_{x,n} \sin (n \phi ), \\ d_y (\phi) &= \sum_{n=1} \lambda_{y,n} \sin (n \phi ), \\  d_z (\phi)  &= \sum_{n=1} \lambda_{z,n} \cos ( n \phi )  .
 \end{aligned}
 \end{equation}
 There is no well-defined winding number for the trajectory of Hamiltonian in phase space. But we find that the $\vec{d}$ always go through the $d_z$ axis when $\phi =0$, $\vec{d} = (0,0, \sum_{n=1} \lambda_{z,n})$ and $\phi = \pi $, $\vec{d} = (0,0, \sum_{n=1}(-1)^n \lambda_{z,n} )$. In the class BDI case, the sign of the multiply of $d_z$ in these two points, $\gamma = \text{sgn} [d_z (\phi = 0)d_z (\phi = \pi)]$ distinguish the odd and even winding number, now it becomes the $\mathbb{Z}_2$ topological index in class D. The trajectory of the $\vec{d}$ vector in class BDI and class D and the corresponding topological invariants are shown in Fig. \ref{WinNumdef}. 
 
 \section{\label{topoinvar} Relation between the winding number and polarization }
 
 According to the AZ topological classification, the topological index in the one-dimension BdG system includes the winding number and two types $\mathbb{Z}_2$ index. The definitions and relations of them are shown here. First is the winding number for class BDI 
\begin{equation}
\nu = \frac{i}{\pi} \int^{2\pi}_{0} dk  \sum_{\alpha \in occ} \langle Su_\alpha (k) \left| \partial_k \right| u_{\alpha} (k ) \rangle.
\end{equation}
where $S$ is the chiral operator, The winding number indicates $\nu = \pm (n_+ - n_- ) $, which means the number of bound states of each chirality at one end of an open chain. The winding number is also called 'skew polarization'  \cite{PhysRevLett.113.046802}. This can be viewed as follows. Note that the Hamiltonian in the chiral symmetry Eq. (\ref{hamchirI}). We could rotate the Hamiltonian into the chiral representation  
\begin{equation}
H = \begin{pmatrix} 0 & h \\ h^\dagger   & 0   \end{pmatrix}.
\end{equation}
To solve this Hamiltonian, we could decompose the eigenstates into two sectors $| \psi_n \rangle = (\psi^A_n , \psi ^B_n)^T$, solve the eigenvalue equation, the two parts have opposite eigenenergies. This is a conclusion of chiral symmetry. 

Using the singular value decomposition for the off-diagonal block $h = U_A \Sigma U^{-1}_B$, in which $U_A$ and $U_{B}$ are both unitary matrices and $\Sigma$ is a diagonal matrix. Note that $U^{-1}_A hh^\dagger U_A = \Sigma^2$, and $U^{-1}_B h^\dagger h U_B = \Sigma^2$, which reveals that $U_A$ and $U_B$ diagonalize $hh^\dagger$ and $h^\dagger h $ respectively, which means the unitary matrices $U_A$ and $U_B$ are the eigenvectors $\psi^A_n$ and $\psi^B_n$. The singular values could be deformed into arbitrary positive values with the same eigenvectors since all singulars are nonzero. Hence it is convenient to set $\Sigma$ to be identity. We denote the flattened Hamiltonian as 
\begin{equation}
Q = \begin{pmatrix} 0 & q \\ q^{-1} & 0 \end{pmatrix}, 
\end{equation}
where $q = U_A U^{-1}_B$ is a unitary matrix and the energy spectrum only takes values of $1$ or $-1$. For the system with translation symmetry, the winding number in the Bloch representation is 
\begin{equation}
\nu = \frac{i}{2\pi} \int^\pi_{-\pi} dk \text{Tr} [q^{-1}(k) \partial_k q(k)],
\end{equation}
since $q(k) = U^A(k) U^{-1}_B(k)$, we could expand it into $\nu = \frac{i}{2\pi} \int^\pi_{-\pi} dk\{ \text{Tr} [U_A^{-1}(k) \partial_k U_A(k)] -  \text{Tr} [U_B^{-1}(k) \partial_k U_B(k)]\}$, with the relation $\text{Tr} (U_q^{-1} \partial U_{q} ) = -\text{Tr} (U_q \partial U^{-1}_{q} ) $ for unitary matrix $U_q$. As mentioned before, the $U_q$ is the eigenvector for the off-diagonal part of Hamiltonian, which means the winding number is the difference of the polarization between the two sublattices, i.e., the meaning for 'skew' polarization. 

At the end of this section, we introduce an important relation between the winding number and the polarization is  
\begin{equation}
2 \gamma = \nu  \mod 2 \>.
\end{equation}
This relation has a rigorous but tedious analytic proven in Ref. \cite{PhysRevLett.113.046802}, while our geometry depiction in Fig. \ref{WinNumdef} gives an intuitive version of it. 

To illustrate this, we use the equivalent topological index in real space since they could avoid the problem of the gauge of the wave function. By replacing the integral and partial to the trace per volume and commutator with the position operator in the open boundary condition, the winding number for class BDI systems in one dimension real space is defined as \cite{PhysRevB.106.045116, PhysRevLett.113.046802}
\begin{equation}
\nu =  \mathcal{T}  \left\{   Q_{BA} \left[ X, Q_{AB} \right]  \right\},
\label{defwni}
\end{equation}
where $\mathcal{T}$ refers to trace per volume, $Q_{AB} = \Gamma_A Q \Gamma_B$ and $\Gamma_\sigma = \sum_{l,\alpha \in \sigma} \left| l,\alpha \rangle \langle l, \alpha \right|$ is the projector on the $\sigma =  A,B$ subspace.  $Q$ is the flatten Hamiltonian in the chiral representation $Q = P_{unocc} - P_{occ}$. Since the winding number defined by trace per volume is not rigorous quantized and susceptible to the system size, which means the definition Eq. (\ref{defwni}) is not algebraic rigorously. 

Alternatively We know the system is topologically equal to the flat Hamiltonian, $Q = P_+ - P_-$, where $P_{\pm}$ are the projection operators for positive or negative eigenstates. Decomposing it by the chiral symmetry $Q = S_+ Q S_- + S_- Q S_+$, one could provide the covariant off-diagonal term $Q_{+-} = S_+ Q S_-$. By recalling $\partial_i = -i [X,\cdot]$, with the position operator $X$ in open boundary condition, The winding number in $2n+1$ dimension real space is \cite{PhysRevLett.113.046802}
\begin{equation}
\nu \!=\! \frac{-(\pi i )^n}{(2n+1)!!} \! \sum_\rho \! (-1)^{\rho} \! \mathcal{T} \! \left\{ \! \prod^{2n+1}_{i=1} \! Q\!_{-+} \left[ X_{\rho i}, Q\!_{+-} \right] \! \right\},
\end{equation}
where $\mathcal{T}$ refers to trace per volume. This formula can be evaluated in translation symmetry breaking situations and independent of the disorder configurations.  

The winding number defined by the trace per volume are sensitive to the scaling of system size. As claimed in the previous section, the winding number is the relative polarization between each sublattice, hence, an alternative equivalent formula to calculate the winding number is proposed as \cite{PhysRevB.106.045116}:
\begin{equation}
\nu = \frac{1}{2\pi i } Tr \left[ \log \left( \chi_A \chi^{-1}_B \right)  \right],
\label{defwnii}
\end{equation}
where $\chi_\sigma = U^{-1}_\sigma \Gamma_\sigma \chi \Gamma_\sigma U_\sigma$ ($\sigma = A, B$) are unitary matrices, $U$ is the eigenvector for the flatten Hamiltonian and $\Gamma_\sigma = \sum_{l,\alpha \in \sigma} |l,\alpha \rangle \langle l,\alpha |  $, this quantity can be considered as the position operator projected onto the $\sigma$ sector of the eigenstate in the occupied band. The quantum mechanical position operator can be chosen as 
\begin{equation}
\chi = \sum_{l,\alpha \in A, \beta = B} e^{i\frac{2\pi}{L}l} \left( | l,\alpha \rangle \langle l,\alpha | + | l ,\beta \rangle \langle l,\beta | \right),
\end{equation}
which could ensure the formula is fitted for the periodic boundary condition, and the winding number defined by this equation could give a strictly quantized number in the half-filling condition. Moreover, this formula enables us to calculate the winding number in any fractional fillings. 

\begin{figure}[tbp]
\centering
\includegraphics[scale=0.26]{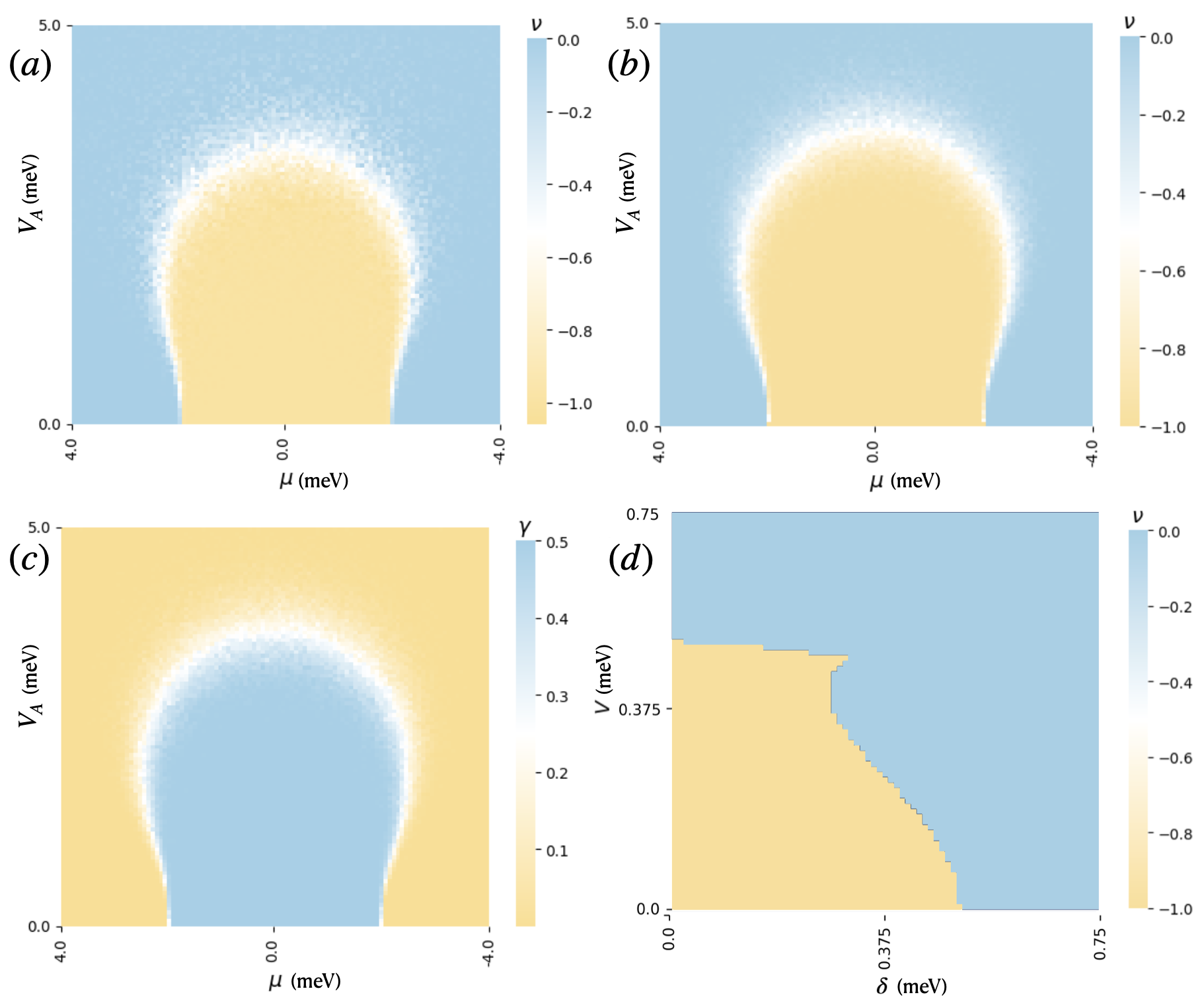}
\caption{ Topological phase diagram for the ordinary Kitaev chain model with Anderson disorder, calculated by Winding number (a) $\nu_1$, (b) $\nu_2$, and (c) polarization $\gamma$, calculated by Eq. (\ref{defwni}), Eq. (\ref{defwnii}) and Eq. (\ref{rspolar}) respectively. Both of them give the same phase boundaries to represent the topological Anderson superconductor states. (d) phase diagram of Hamiltonian Eq. (\ref{HdimerizedBDI}) with the quasiperiodic potential. The region of topology has $\nu = -1$ and $\gamma = 1/2$, otherwise the trivial phase obtain $\nu = \gamma = 0$. We select the system size $N = 1597$.  }
\label{testwnpolar}
\end{figure}

In practice, we draw the phase diagram of Anderson disorder $V_A$ manipulated ordinary Kitaev chain model $H = - \sum_j \left( tc^\dagger_{j+1} c_j +\Delta c^\dagger_{j+1} c^\dagger_j - V_A f(j) c^\dagger_j c_j \right) +h.c. $, in Fig. \ref{testwnpolar}, by calculating the winding number, defined by (a) Eq. (\ref{defwni}), (b) Eq. (\ref{defwnii})  and (c) polarization, Eq. (\ref{rspolar}) as a comparison. The $\nu = 1$, $\gamma = 1/2$ indicates the topological phase and $\nu = 0$, $\gamma = 0$ corresponding to the trivial phase. The earlier discussions have told us that the topological phase transition points are $|\mu|/t = 2$ at the clean $V = 0$ case. However, as the enhance of disorder strength, the topological phase undergoes significant inflation, and the maximum point of the phase boundary expands to about $|\mu| / t \sim 2.7$, which means the system enters into the topological Anderson superconductor phase. All of these phase diagrams calculated by the three invariants obtains the same results, which validates the correctness of our method.

\begin{figure*}[htbp]
\centering
\includegraphics[scale=0.43]{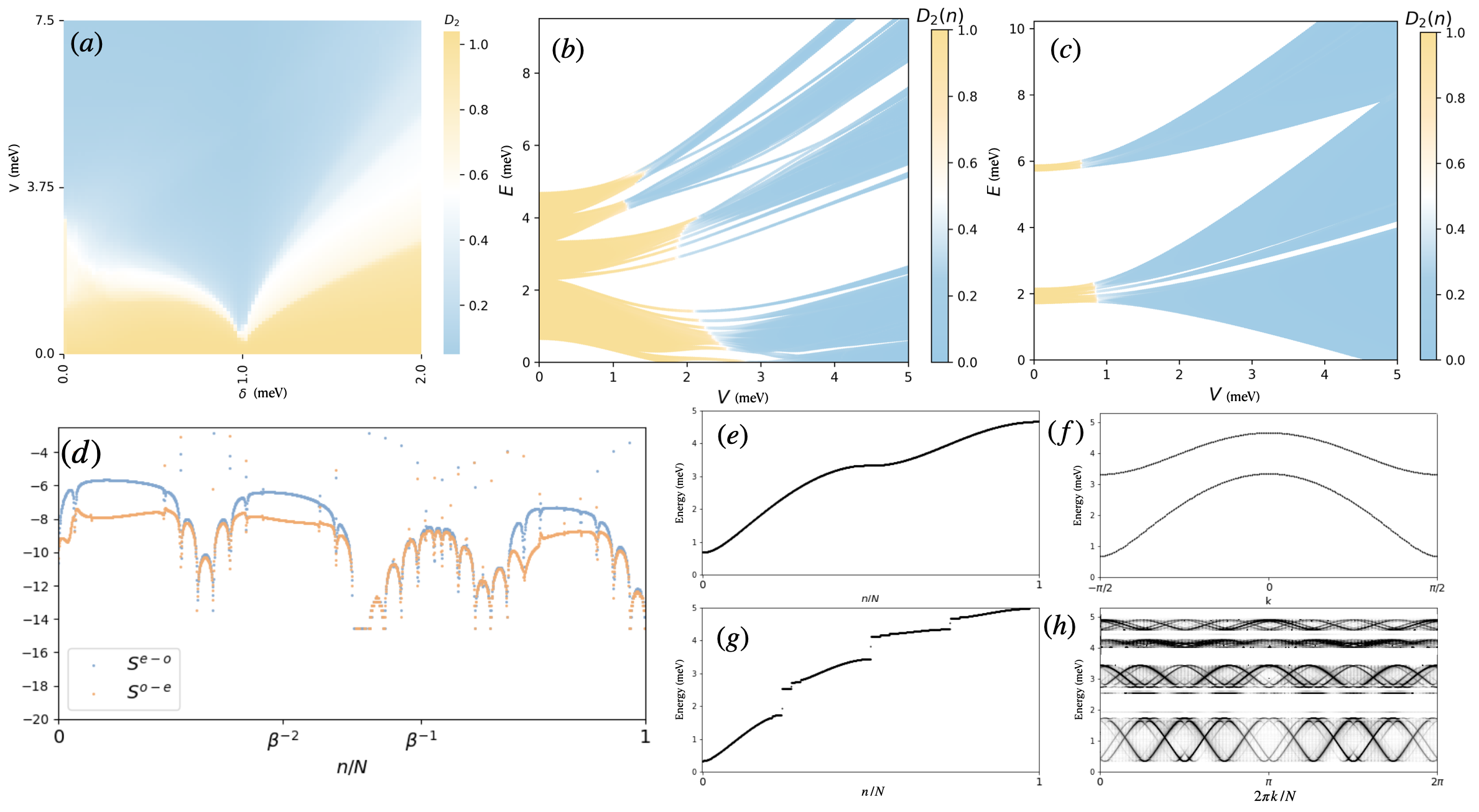}
\caption{  (a) Phase diagram for the dimerized model, Hamiltonian in Eq. (\ref{HdimerizedBDI}), calculated by the mean fractal dimension $D_2$ with the system size $N = 1597$. Energy spectrum with the dimerization strength (b) $\delta = 0.34$ and (c) $\delta = 0.95$ in terms of the quasiperiodic potential $V/t$. The color map indicates the fractal dimension $D_2(n)$ of each eigenstate. (d) The logarithm of the even-odd (odd-even) energy level spacings in black (red) for the Hamiltonian with quasiperiodic potential $V = 1.34$. The corresponding real space energy spectrum is shown in (e) clean case $V = 0$ and (g) disorder case $V = 1.34$, we also draw them in (f) Bloch state for clean case and (h) discrete Fourier transformation for disorder case. Here we select the size of system $N = 4181$ except the phase diagram. }
\label{multifractal}
\end{figure*}

\section{\label{InterpreteIPRNPR}Interpretation of the Inverse and Normalized Participation Ratios}

To understand the relationship between the inverse/normalized participation ratio (IPR/NPR) and localization behavior, we begin with the case of an ergodic state. According to Bloch's theorem, in a periodic potential where $\phi(x) = \phi(x + a)$, the wave functions are delocalized across the entire lattice. In lattice models with $L$ sites and periodic boundary conditions (effectively forming a ring), a fully extended state has equal amplitude on each site. For such a state, the normalization condition gives $\phi_i \propto 1/\sqrt{L}$. Using the definition of the IPR, $I_x = \sum_{i=1}^L |\phi_i|^4 \propto {1}/{L}$, which clearly tends to zero as $L$ increases, with a scaling exponent of $-1$. In contrast, for a fully localized state, e.g., one that is nonzero only at a single site, we have $\phi_i = \delta_{i,i_0}$, and hence $I_x \propto 1 $ independent of $L$.

The behavior of the NPR, defined in momentum space, complements the above analysis. Due to the Fourier transform, the wave function of an ergodic state in real space becomes sharply peaked, i.e., localized, in momentum space. As a result, for an extended state, $I_x^{(j)} \sim L^{-1}, \quad I_k^{(j)} \sim \mathcal{O}(1)$, while for a localized state, $I_x^{(j)} \sim \mathcal{O}(1), \quad I_k^{(j)} \sim L^{-1}$. For intermediate, or multifractal, states, which are neither fully extended nor fully localized, both $I_x$ and $N_k$ take finite values that scale nontrivially with system size. To characterize this regime, we could define the $\eta$-quantity $\eta = \log_{10} \left( \bar{I}_x \times \bar{I}_k \right)$ where $\bar{I}_x$ ($\bar{I}k$) denote averaged IPR (NPR) over the ensemble. In either the fully extended or fully localized limit, one of the two terms in the product scales as $\mathcal{O}(L^{-1})$, leading to $\eta \leq  -\log_{10}L$. In contrast, in the intermediate phase, both $\bar{I}_x$ and $\bar{I}_k$ remain finite ($\sim \mathcal{O}(1)$), resulting in $\eta \sim \mathcal{O}(0)$, independent of $L$.

\section{\label{centrotsc} Disorder manipulate dimerized centrosymmetric topological superconductors}

In this section, we consider the sublattice symmetry-breaking model but keep the time-reversal symmetry. The Hamiltonian of such a system is represented by 
\begin{equation}
\begin{aligned}
\mathcal{H}  \!&= \! - \sum_j  \left[ (2+ \delta) c^\dagger_{A,j} c_{B,j}  +(2- \delta)  c^\dagger_{B,j}c_{A,j+1}   \right]  \\ & \quad+  \sum_j \left[  (1+\delta)  c^\dagger_{A,j} c^\dagger_{B,j}+ (1- \delta) c^\dagger_{B,j} c^\dagger_{A,j+1}  \right] \>\\ & \quad - \sum_j V \left[   f_{2j} c^\dagger_{A,j} c_{A,j}  +   f_{2j+1}  c^\dagger_{B,j} c_{B,j}  \right] +h.c. \>,
\end{aligned}
\label{HdimerizedBDI}
\end{equation}
where we select the hoppings $t_1,t_2 = 2\pm \delta$ and pairings $\Delta_1, \Delta_2 = 1\pm \delta$, these kind of parameters make sure the Hamiltonian still hold the time-reversal and inversion symmetry. The quasiperiodic potential selected as $f_i = 2 \cos (2\pi \beta i +\phi)$ for each site $i$. In Fig. \ref{multifractal} (a), we obtain the global phase diagram with the mean fractal dimension Eq. (\ref{avefracdim}), a clear mobility edge separating ergodic states ($D_2 = 1$) from the multifractal states ($D \sim 0.4$). We draw the energy-dependent transition between ergodic-multifractal states and ergodic-localized states in Fig. \ref{multifractal} (b) and (c), which is determined by Eq. (\ref{deffracdim}), the color map indicates the fractal dimension of the state corresponding to each energy $E/t$. A mobility edge separating ergodic states from localized states ($D_2 = 0$). For each band, there is a sharp transition at a different critical value of the quasiperiodic potential $V$. 

Calculate the energy level spacing, Eq. (\ref{defels}) in Fig. \ref{multifractal} (d) for the Hamiltonian Eq. (\ref{HdimerizedBDI}). The ergodic region of the spectrum is located at the low energy region with a clear gap. In contrast, we expect to see no difference between the even-odd (odd-even) cases for localized states. Furthermore, the multifractal regions will show strongly scattered distributions for both of them. This approach not only provides a statistical description of phase transitions but also helps in understanding the complex quantum states and localization phenomena in quasiperiodic potential systems.

We noticed that the blocklike structure does not rigorously hold in our model. In Fig. \ref{multifractal} (b), the eigenstates in the block with the second highest energy entered the localized state before the eigenstates in the block with the first highest energy. This kind of anomalous mobility edge behavior originates from the band hybridization brought by our unique SSH order. The inversion symmetry ensures the double degeneration of the energy bands, which makes the logarithm of $S^{o-e}$ and $S^{e-o}$ distinguishable in the ergodic phase. However, the nonzero SSH order breaks the degeneration of the folded bands which makes an indirect gap. We plot the energy bands without the disorder in real space and momentum space, view in Fig. \ref{multifractal} (e) and (f), which reveals the large density of states in the overlapping band region. When we enhance the disorder strength, the energy gaps are opened between each block structure, shown in Fig. \ref{multifractal} (g). 

Although, we could not directly draw the energy bands in the momentum space since the translation periodic is infinity. In finite size, we could use the discrete Fourier transformation $\phi^{(j)}_k (l) = \frac{1}{L} \sum^L_{p=1} e^{-i2\pi pl/L} \phi^{(j)}_x (p)$ to calculate the energy spectrum for each eigenstates in Fig. \ref{multifractal} (h). The quantum number is $2\pi k /N$, $k = 0,1,..., N-1$, and the color map indicates the modulus of the wave function. According to the results, we find that in the hybridization region, the stronger inter-band scattering disrupts electron coherence and enhances disorder effects, making electrons in this region more prone to localization. This phenomenon in our system highlights the complexity of localization behavior in multi-band systems and offers new insights into understanding mobility edges, the robustness of topological states, quantum transport, and their applications in topological quantum devices.

Furthermore, we draw the winding number of the Hamiltonian Eq. (\ref{HdimerizedBDI}) with the quasiperiodic disorder in Fig. \ref{testwnpolar} (d), which exhibits the similar topological Anderson superconductor phases with clear phase boundary. Combine with Fig. \ref{multifractal} (a), one could obtain the universal phase diagram which contains information on both the localization features and topological properties. The disorder-induced topological state is an exciting discovery because it challenges the traditional view that topological phases are solely determined by the intrinsic symmetries of the system, independent of disorder.

\section{\label{SupercurrentindSC} Supercurrent induced superconductivity}

The underlying mechanism is the finite center-of-mass momentum of Cooper pairs via the minimal coupling substitution $\textbf{k} \to \textbf{k} - {e} \textbf{A}(r) /{\hbar}$, which arises through the Meissner effect and manifests as a spatially varying phase in the superconducting order parameter \cite{ WOS:000800045000012, WOS:000840612900002, PhysRevB.108.174516, WOS:000985894000001}.
\begin{equation}
\Delta (\textbf{r}) = \Delta\exp(i\frac{2e}{\hbar} \textbf{A} \cdot \textbf{r}),
\end{equation}
 at position $\textbf{r} $. We can rewrite the coefficient as finite momentum $\textbf{q} = 2e\textbf{A} /{\hbar} $, where $\textbf{q}$ in the direction of the screening current that is perpendicular to the extra supercurrent.  
 
 Make variable substitution 
 \begin{equation}
 \begin{aligned}
  \Delta \exp(i \textbf{q} \cdot \textbf{r} ) &= \Delta \exp(i \frac{\textbf{q}}{2} (2 \textbf{r} + \bold{a} ) - i \frac{\textbf{q}}{2} \bold{a} )  
 \end{aligned}
 \end{equation}
 where we set $\phi = \frac{\textbf{q}}{2} $ and $\bold{a}$ is the lattice constant, at a fixed site $\textbf{r}$ in $r_j$, $\textbf{r} + \bold{a}$ in next lattice site $r_{j+1}$, we can write the pairing as
 \begin{equation}
  \Delta \exp(i \textbf{q} \cdot \textbf{r} ) = \Delta \exp(i \phi (r_{j+1} + r_j )  ).
 \end{equation}
 After neglecting the last no spatial varying phase term $ \exp( - i {\textbf{q}} \bold{a}/{2} ) $. The superconductor pairing in external supercurrent becomes 
 \begin{equation}
 \Delta (\textbf{r}) = e^{i\phi \left( r_{j+1} +r_j \right)} | \Delta | .
\end{equation}
We begin from a {complex} fermion $c_j$ at each lattice site $j$, where we can modulate the staggered hopping amplitudes between lattice sites, by adjusting the lateral stress applied within the substrate. The lattice BdG Hamiltonian for the 1-D wire with dimerized $p$-wave superconductivity \cite{PhysRevB.90.014505}, $H = H_0+H_i$, where the free fermion part: 
\begin{equation}
\begin{aligned}
H_0 &= -\sum_j [ t_1 c^\dagger_{A,j} c_{B,j}  + t_2 c^\dagger_{B,j}c_{A,j+1}   +h.c. ]  \\ & \quad  + \sum_j [ \mu_1 c^\dagger_{A,j} c_{A,j}  +  \mu_2  c^\dagger_{B,j} c_{B,j}  ]  ,
\end{aligned}
\end{equation} 
and the finite momentum pairing terms 
\begin{equation}
\begin{aligned}
H_i  & = \sum_j [  e^{i\phi_1(r_{A,j}+r_{B,j})}  |\Delta_1| c^\dagger_{A,j} c^\dagger_{B,j} \\ & \qquad +e^{i\phi_2(r_{B,j}+r_{A,j+1})} |\Delta_2|  c^\dagger_{B,j} c^\dagger_{A,j+1}  ] + h.c.\>  . 
\end{aligned}
\end{equation} 
Make a substitution, and gauge the phase into the operator 
\begin{equation}
\begin{aligned}
c^\dagger_{\alpha,j} \!  \to  \!e^{-i \phi_\alpha r_j} c^\dagger_{\alpha,j}, \qquad   c_{\alpha,j}  \! \to \!e^{+i \phi_\alpha r_j} c_{\alpha,j} , 
\end{aligned}
\end{equation}
where $\alpha = A, \> B$. 
Then the hopping terms of the Hamiltonian becomes 
\begin{equation}
\begin{aligned}
H_h \! &= -\sum_j [ t_1 e^{i\phi_1 (r_{A,j} - r_{B,j})} c^\dagger_{A,j} c_{B,j}  +h.c. ]  \\ & \quad - \sum_j [ t_2 e^{i\phi_2 (r_{B,j} - r_{A,j+1})} c^\dagger_{B,j}c_{A,j+1}   +h.c. ] ,
\end{aligned}
\end{equation}
and the pairing terms become
\begin{equation}
H_p = \sum_j [   |\Delta_1| c^\dagger_{A,j} c^\dagger_{B,j}+ |\Delta_2|  c^\dagger_{B,j} c^\dagger_{A,j+1} + h.c. ].
\end{equation}
  Since the intra-cell and inter-cell have different screen currents and intervals, we could construct the effective staggered complex hopping amplitude and pairings where we select the 
 \begin{equation}
 \begin{aligned}
 \phi_1 &= \phi_1 (r_{A,j} - r_{B,j}),  \\
 \phi_2 &= \phi_2 (r_{B,j} - r_{A,j+1}).
 \end{aligned}
 \end{equation}
 After combining with the chemical potential 
 \begin{equation}
 H_c =  \sum_j [ \mu_1 c^\dagger_{A,j} c_{A,j}  +  \mu_2  c^\dagger_{B,j} c_{B,j}  ] ,
 \end{equation}
 one can reproduce the Eq. (\ref{ham0}) in section \ref{Model}.

\bibliography{TopoSupcon}

\end{document}